\documentclass[
reprint,
superscriptaddress,
altaffilletter,
nofootinbib,
amsmath,
amssymb,
aps,
prd,
]{revtex4-1}

\usepackage{graphicx}
\usepackage{hyperref}
\usepackage{booktabs}
\usepackage{aas_macros} 
\usepackage{orcidlink}
\usepackage{siunitx}
\usepackage[normalem]{ulem}
\usepackage[caption=false]{subfig}

\graphicspath{{figures/}}

\newcommand{\DocumentID}{2400426}
 
\usepackage{xcolor} 

\newcommand{\dcc}{LIGO-P}
\renewcommand{\log}{\ln}

\begin{document}

\title{Mitigating the impact of noise transients in gravitational-wave searches using reduced basis timeseries and convolutional neural networks}

\author{Ryan Magee \orcidlink{0000-0001-9769-531X}}
\email{rmmagee@caltech.edu}
\affiliation{LIGO, California Institute of Technology, Pasadena, CA 91125, USA}

\author{Ritwik Sharma \orcidlink{0000-0003-1858-473X}}
\affiliation{Institut f{\"u}r Physik und Astronomie, Universit{\"a}t Potsdam,
Haus 28, Karl-Liebknecht-Str. 24/25, 14476, Potsdam, Germany}

\author{Ananya Agrawal}
\affiliation{Blue Valley West Highschool, Overland Park, KS 66085, USA}

\author{Rhiannon Udall \orcidlink{0000-0001-6877-3278}}
\affiliation{LIGO, California Institute of Technology, Pasadena, CA 91125, USA}

\begin{abstract}

Gravitational-wave detection pipelines have helped to identify over one hundred
compact binary mergers in the data collected by the Advanced LIGO and Advanced
Virgo interferometers, whose sensitivity has provided unprecedented access to
the workings of the gravitational universe. The detectors are, however, subject
to a wide variety of noise transients (or glitches) that can contaminate the
data. Although detection pipelines utilize a variety of noise mitigation
techniques, glitches can occasionally bypass these checks and produce false
positives. One class of mitigation techniques is the signal consistency check,
which aims to quantify how similar the observed data is to the expected signal.
In this work, we describe a new signal consistency check that utilizes a set of
bases that spans the gravitational-wave signal space and convolutional neural
networks (CNN) to probabilistically identify glitches. We recast the basis
response as a grayscale image, and train a CNN to distinguish between
gravitational-waves and glitches with similar morphologies. We find that the
CNN accurately classifies $\gtrsim 99\%$ of the responses it is shown. We
compare these results to a toy detection pipeline, finding that the two methods produce similar false
positive rates, but that the CNN has a significantly higher true positive rate.
We modify our toy model detection pipeline and demonstrate that
including information from the network increases the toy pipeline's true positive rate by $4-7\%$ while decreasing the false positive rate to a data-limited bound of $\lesssim 0.1\%$.
\end{abstract}

\maketitle

\section{Introduction}

The Advanced LIGO~\cite{LIGOScientific:2014pky} and Advanced
Virgo~\cite{VIRGO:2014yos} interferometers have provided a new way to observe
our Universe. Since the detectors began observations in 2015, the
LIGO-Virgo-Kagra collaboration (LVK)~\cite{LIGOScientific:2021djp,LIGOScientific:2021usb,LIGOScientific:2020ibl,LIGOScientific:2018mvr} and other
groups~\cite{Nitz:2018imz,Magee:2019vmb,Venumadhav:2019lyq,Zackay:2019btq,Nitz:2020oeq,Nitz:2021uxj,Nitz:2021zwj,Olsen:2022pin}
have announced the detection of gravitational waves (GWs) from over 100 unique
compact binary mergers. In aggregate, these detections have shed light on the
mass distribution of black holes in binaries~\cite{KAGRA:2021duu}, facilitated
tests of general
relativity~\cite{LIGOScientific:2018dkp,LIGOScientific:2021sio}, and probed nuclear
physics~\cite{LIGOScientific:2018cki,LIGOScientific:2020aai}.
Individually, they have challenged our understanding of stellar
evolution~\cite{LIGOScientific:2020zkf,LIGOScientific:2020iuh,LIGOScientific:2024elc}
and yielded insight into the formation of compact binaries~\cite{Zevin:2020gbd}.

There are, however, a number
of selection effects introduced by GW searches that can
reduce our sensitivity to astrophysical sources and impact our understanding of
the Universe. Excluding sources that are
selected against can have a profound impact on downstream
analyses~\cite{Messenger:2012jy,Magee:2023muf}. Some
selection effects are introduced directly by the search; for example, the
extent of the searched parameter space is set in advance for modeled searches.
Many selection biases, however, are unknown in advance and must be mitigated
against during the detection process. Most notably, this includes the impact of non-stationary noise
in the detector.

Non-Gaussian noise transients, or \emph{glitches}, pose a dual threat to
detection pipelines. Glitches can contaminate the background of the search,
causing the pipeline to assign systematically lower significances to
astrophysical signals~\cite{LIGOScientific:2017tza}. Perhaps more dangerously, however, glitches can masquerade as
spurious foreground signals and contaminate the purity of catalogs or
low-latency GW alerts.  When multiple interferometers are simultaneously
collecting data, search pipelines can mandate that candidates are observed across data streams.  Enforcing coincidence heavily suppresses the background; although
transient noise can mimic GWs, there is a very low chance that similar deviations from Gaussianity are observed
simultaneously in multiple detectors.  If only one data stream is available,
however, pipelines rely heavily on other quantities incorporated in their complex
ranking statistics~\cite{Tsukada:2023edh,Aubin:2020goo,Nitz:2018rgo} to
separate signal from noise. Unfortunately, sufficiently loud transients or
those that exhibit morphologies uncharacteristic of previous data can still
produce false positives.

Machine learning has recently emerged as a promising avenue towards
understanding and characterizing detector noise. Machine learning-based
approaches to GW
astrophysics have exploded in recent
years (see~\cite{Cuoco:2020ogp,Stergioulas:2024jgk} for reviews)~\cite{George:2017pmj,Gabbard:2017lja,Nousi:2022dwh,Marx:2024wjt,Alfaidi:2024ioo,Schafer:2022dxv,Mishra:2022ott,Krastev:2019koe,Schafer:2020kor,Wei:2020sfz,Baltus:2021nme,Yu:2021vvm,Baltus:2022pep,Astone:2018uge,Antelis:2021qak,LopezPortilla:2020odz,Gabbard:2019rde,Dax:2021tsq,Chatterjee:2022ggk,Chatterjee:2022dik,Dax:2022pxd,Wong:2023lgb,Schmidt:2020yuu,Chua:2018woh,Khan:2020fso,Tissino:2022thn,Thomas:2022rmc,Magee:2024yal},
largely due to their ability to model complex, non-linear systems. Within GW data
analysis, it promises to be especially useful in understanding the behavior of
unmodeled systems, such as the transient noise present in all GW
interferometers. Deep neural networks, binary classifiers, and convolutional neural networks (CNNs) have all found use in removing and classifying detector noise~\cite{Vajente:2019ycy,Essick:2020qpo,Saleem:2023hcm}.  Perhaps
the most well known application of machine learning to detector noise is
\texttt{GravitySpy}~\cite{Zevin:2016qwy,Glanzer:2022avx}, which has successfully classified
thousands of glitches across a wide range of morphologies and provided reliable
training data for other glitch classification and generation
pipelines~\cite{Merritt:2021orq,2022PhRvD.106b3027L,2022arXiv220509204L}.
\texttt{GravitySpy} demonstrated that different classes of noise transients
imprint unique signatures in the raw GW data that are identifiable by CNNs.  In
this work, we ask a related question: does the projection of glitches onto the
compact binary coalescence (CBC) signal space elicit a unique response that can
be leveraged to improve detection pipeline robustness against noise?

\begin{figure*}
\centering
\subfloat[]{%
        \includegraphics[width=0.3\linewidth]{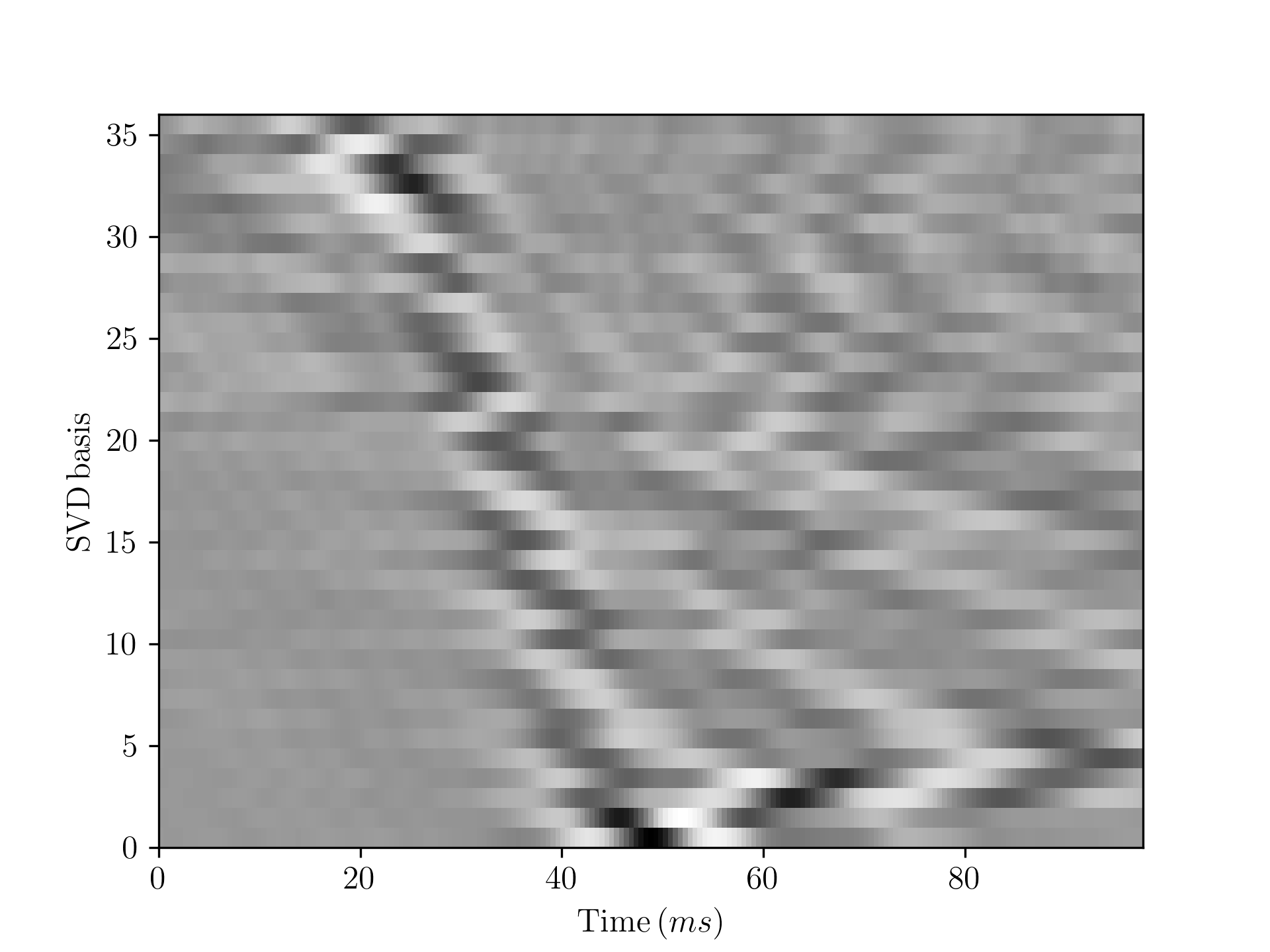}
        \label{fig:signal_t}
}
\subfloat[]{%
        \includegraphics[width=0.3\linewidth]{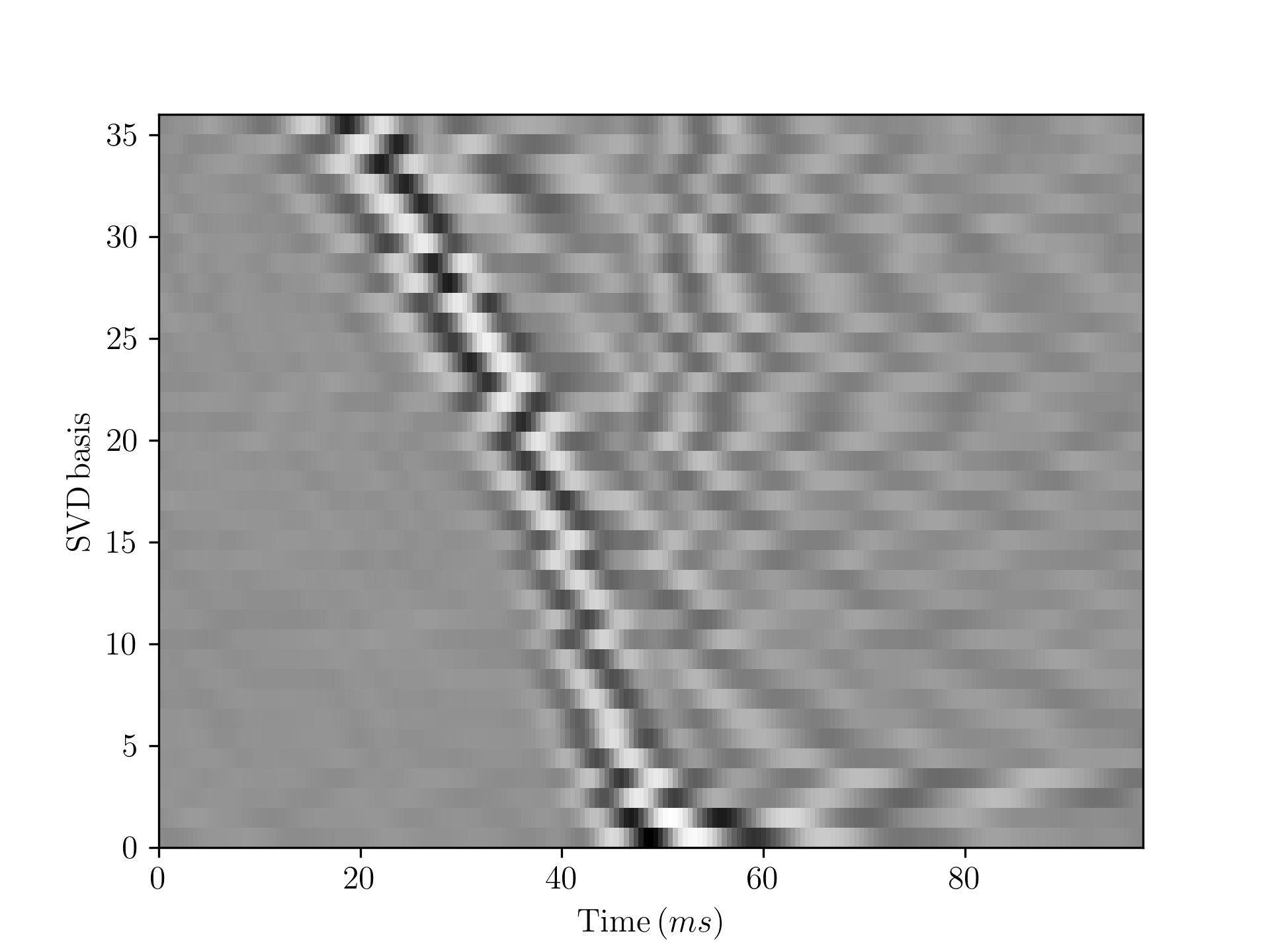}
        \label{fig:blip_t}
}
\subfloat[]{%
        \includegraphics[width=0.3\linewidth]{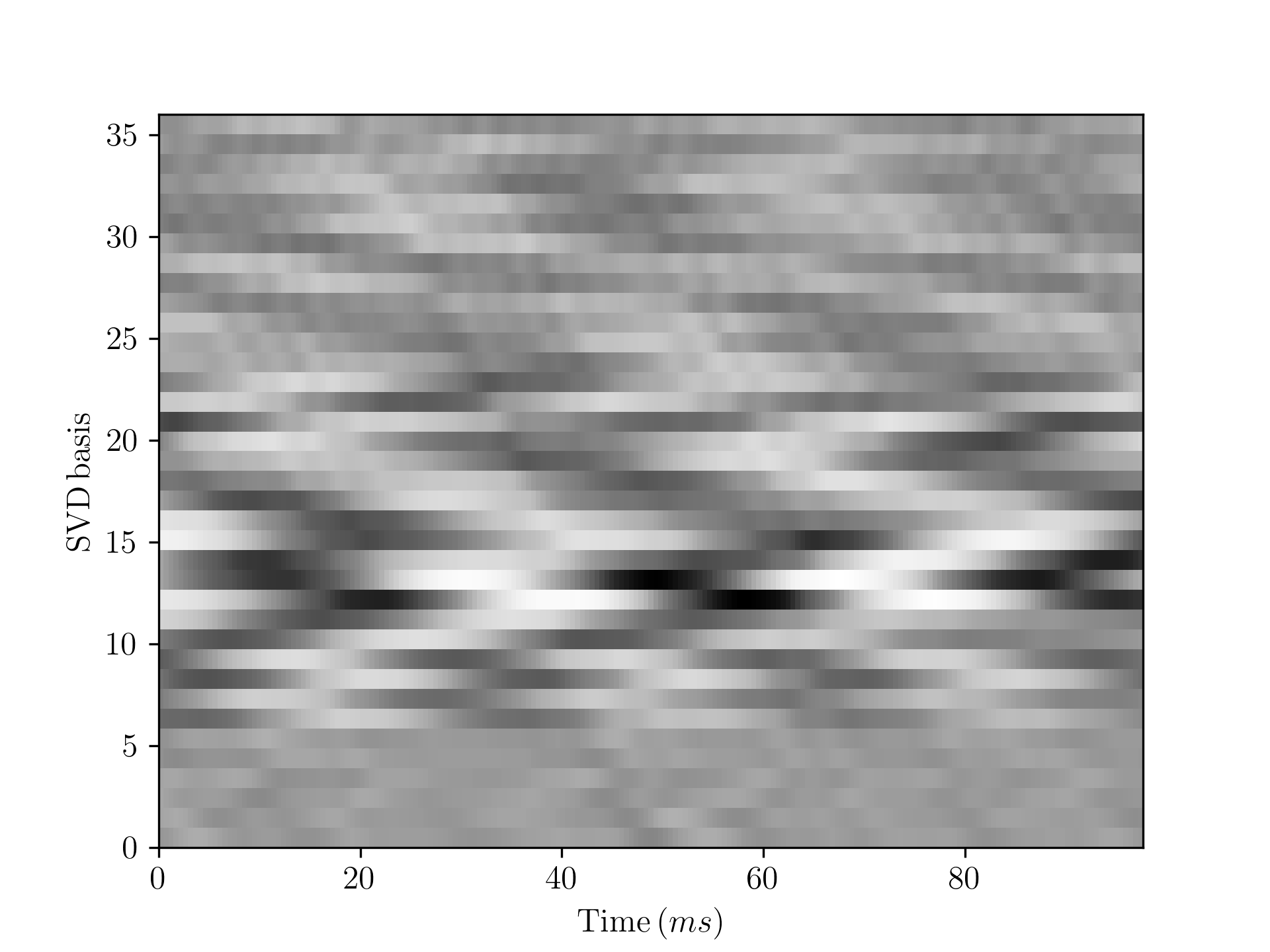}
        \label{fig:arch_t}
}\\
\caption{\label{fig:t_domain}The basis vector response in the time domain to a simulated GW~\ref{fig:signal_t}, blip glitch~\ref{fig:blip_t}, and scattering
arch~\ref{fig:arch_t}. The color of each pixel corresponds to the SNR measured
in that basis, at that time. Darker values correspond to higher SNRs.}
\end{figure*}

\begin{figure*} 
\centering
\subfloat[]{%
        \includegraphics[width=0.3\linewidth]{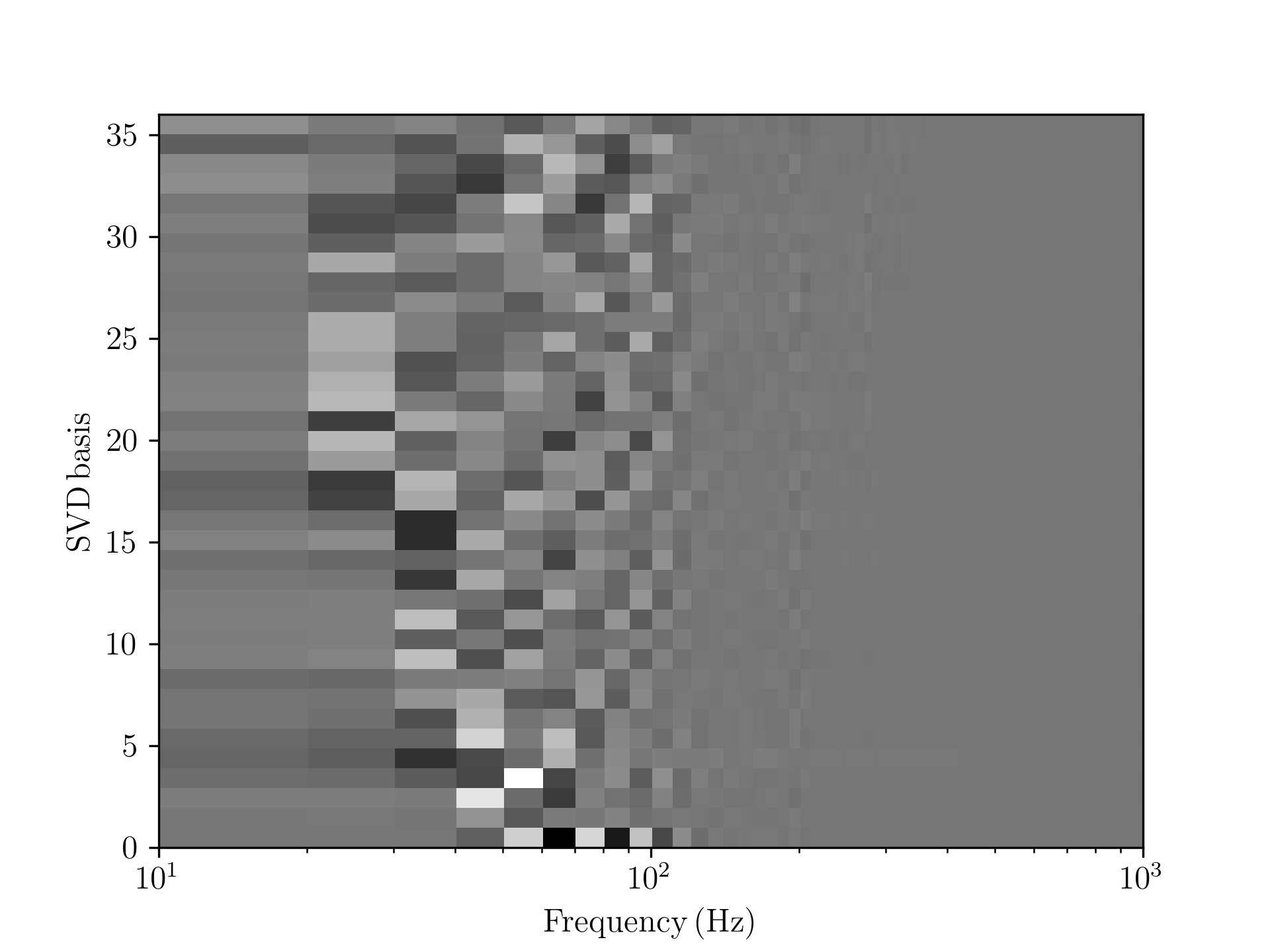}
        \label{fig:signal_f}
}
\subfloat[]{%
        \includegraphics[width=0.3\linewidth]{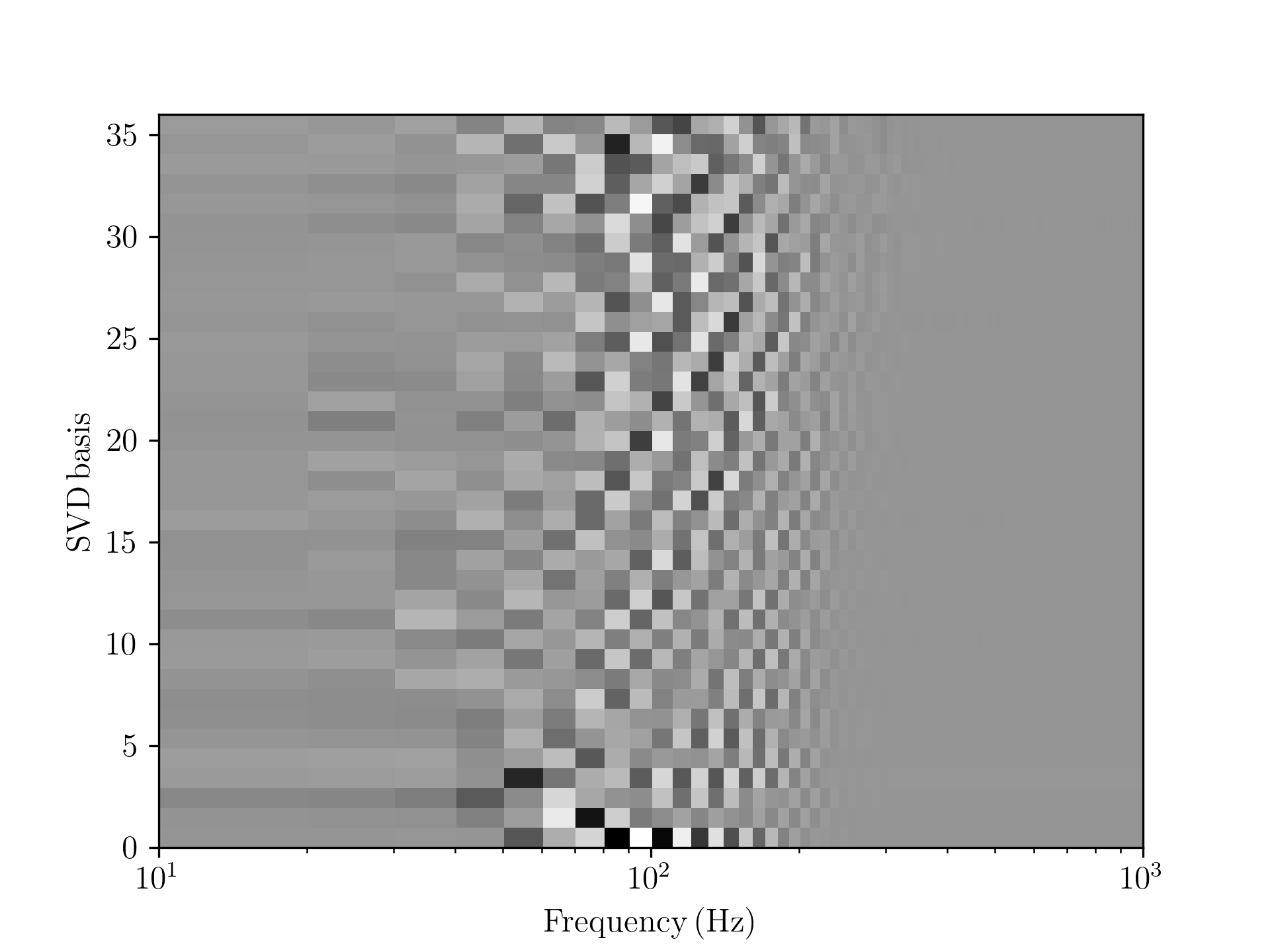}
        \label{fig:blip_f}
}
\subfloat[]{%
        \includegraphics[width=0.3\linewidth]{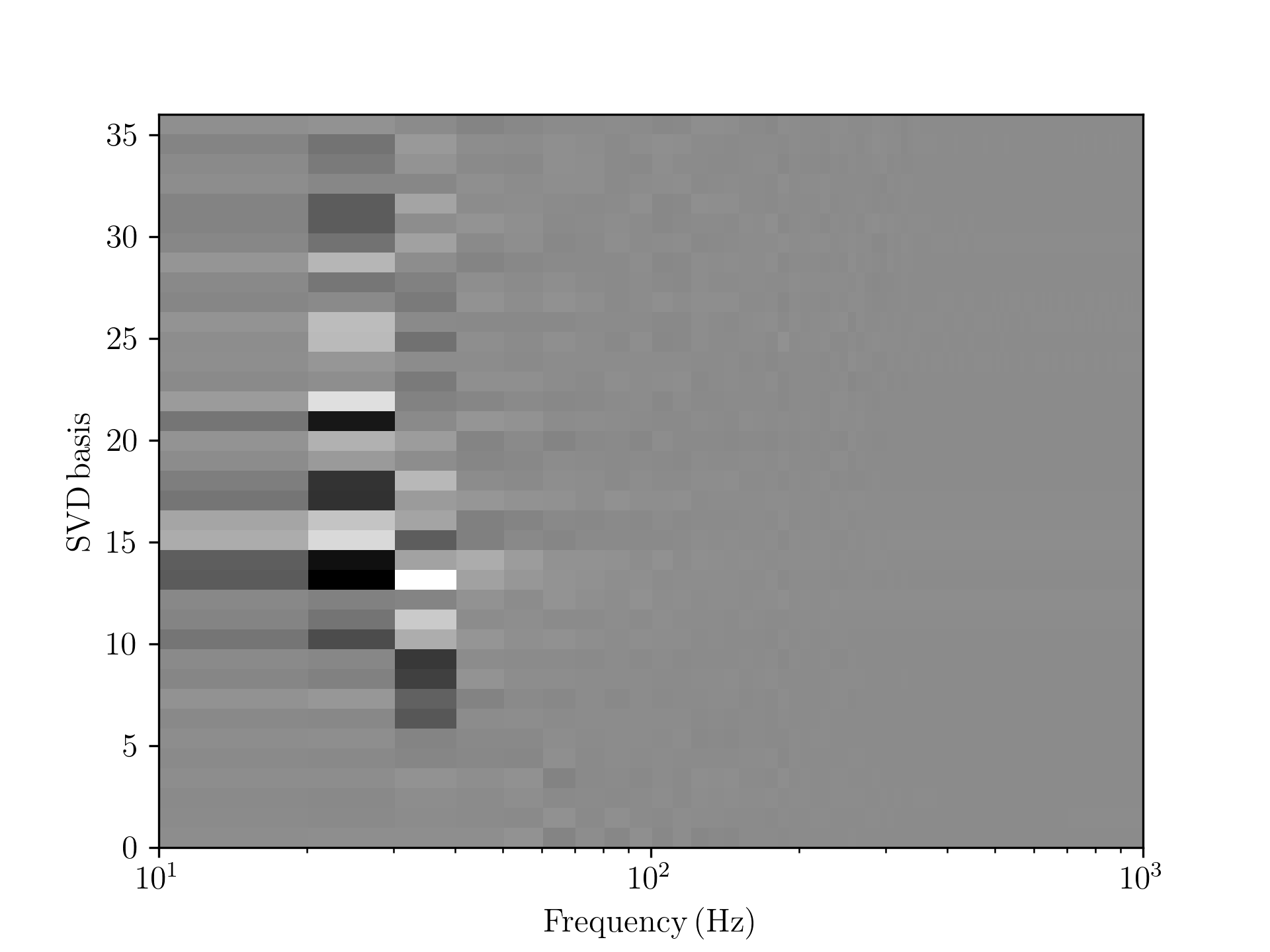}
        \label{fig:arch_f}
}
\caption{\label{fig:f_domain}The basis vector response in the frequency domain
to a simulated GW~\ref{fig:signal_f}, blip glitch~\ref{fig:blip_f}, and scattering
arch~\ref{fig:arch_f}. The color of each pixel corresponds to the SNR measured
in that basis, at that frequency. Darker values correspond to higher SNRs. For
compactness, we only plot the positive frequency components of the Fourier
Transform.}
\end{figure*}

Here, we describe a new, machine learning-based approach to identifying and mitigating the impact
of noise transients --- specifically, blip glitches~\cite{Cabero:2019orq} --- in detection pipelines.
We focus on regions of the GW signal space that either 1) contain candidate CBCs frequently
mischaracterized as noise transients~\cite{Davis:2020nyf} or 2) are
representative of typical LVK detections. 
We recast information presently collected but discarded by 
the GstLAL-based inspiral pipeline~\cite{Messick:2016aqy,Sachdev:2019vvd,2021SoftX..1400680C,Tsukada:2023edh} as grayscale images and use 
CNNs~\cite{LeCun:2015a,2014arXiv1404.7828S} to assign a glitch probability to
GW candidates in local regions of the GW parameter space. In
section~\ref{sec:methods}, we provide background on the
search pipeline and the glitch models we use. In section~\ref{sec:study}, we
describe our specific application of CNNs to detection pipeline outputs. We
describe the study we carry out and the training data that we generate. In
section~\ref{sec:results}, we present our results and compare them to a
toy model detection pipeline. We propose a modification to the toy pipeline that incorporates information from the CNN and demonstrate that it is robust against blip glitches. Finally, in~\ref{sec:conclusions}
we discuss our results and present prospects for the
future.

\section{Methods and motivation}
\label{sec:methods}
\subsection{GstLAL}
\label{sec:gstlal}

Matched filter-based pipelines correlate the expected gravitational wave
emission for a CBC, or \emph{template}, with the data to calculate the
complex-valued signal-to-noise ratio (SNR):
\begin{equation}
z_j(t) = \langle \Re(h_j(t)) | d(t) \rangle + i \langle \Im(h_j(t)) | d(t) \rangle
\end{equation}
where $\langle \cdot | \cdot \rangle$ denotes the noise-weighted inner product,
\begin{equation}
\langle a | b \rangle \equiv 4 \Re \int_0^{\infty}
\mathrm{d}f\frac{a^*(f)b(f)}{S_n(f)} \end{equation}
where $h_j(t)$ is the complex-valued template and $S_n(f)$ denotes the one-sided power spectral density (PSD).  Detection pipeline compute $\mathcal{O}(10^6)$ SNRs at a time to minimize loss of sensitivity to astrophysical transients. To reduce the cost of matched filtering, the
GstLAL-based inspiral
pipeline~\cite{Messick:2016aqy,Sachdev:2019vvd,2021SoftX..1400680C,Tsukada:2023edh}
utilizes singular value decomposition (SVD) to identify orthonormal bases for
local regions of the parameter space~\cite{Cannon:2010qh}. Under this
formalism, the measured SNR for template $j$ can be written as
\begin{equation}
z_j(t) = \sum_\mu a_j^\mu \, \langle u_\mu(t) | d(t) \rangle
\end{equation}
where $a_i^\mu$ is the complex valued reconstruction coefficient associated
with template $j$ for basis vector $u_\mu(t)$.

In stationary, Gaussian noise, the SNR is the optimal detection
statistic. Detector data, however, are neither stationary nor Gaussian, so
pipelines instead rely on complex ranking statistics that incorporate a number
of noise mitigation techniques. These include signal consistency checks such as
$\xi^2$, an autocorrelation-based quantity~\cite{Messick:2016aqy} that describes a single template's
consistency in a small region of time near a candidate's peak SNR:
\begin{equation} 
\xi_j^2 = \frac{\int_{-\delta t}^{\delta t}\mathrm{d}t \, | z_j(t) - z_j(0)\, R_j(t) |^2 \,}{\int_{-\delta t}^{\delta t}\mathrm{d}t \, (2 - 2 \left| R_j(t) \right|^2) \,}\,.
\label{eq:consistency-def}
\end{equation}
Here, $R_j$ represents the
autocorrelation for template $j$, and the integrand in the denominator is a
normalization term that arises from the expectation value of this quantity in
Gaussian noise~\cite{Messick:2016aqy}. GstLAL records the $\xi^2$ and
$\rho \equiv | z |$ for non-coincident\footnote{Background is collected only during times at which two or more interferometers of comparable sensitivity are observing.}, background triggers in a histogram, and subsequently uses
kernel density estimation to smooth the histogram and estimate the associated
probability densities for a local group of templates $\theta$: $P(\xi^2, \rho\,
| \theta, \mathrm{noise})$.
Glitches that cause a template to measure $\xi^2/\rho^2 \lesssim 0.01$ are assigned a small value of $P(\xi^2/\rho^2,
\rho\, | \theta, \mathrm{noise})$ and a high significance. 

Here, we aim to augment our knowledge of the data surrounding a candidate by
leveraging measurements made by the search across all local templates, that is, 
the set of raw basis responses:
\begin{equation}
Q_\mu(t) \equiv \langle u_\mu(t) | d(t) \rangle \, .
\end{equation} 
There is numerical evidence that the bases produced by SVD are
complete~\cite{Cannon:2011xk} and can be interpolated to accurately reconstruct arbitrary waveforms~\cite{Cannon:2011rj,Magee:2024yal}. In aggregate, they span the space of CBC
waveforms and may provide more information than any single template on its own.
Previous work has successfully used $Q_{\mu}(t)$ to define an alternative detection
statistic for compact binary coalescences~\cite{Cannon:2011tb}. In this work,
we instead use them to assign a glitch probability, $p_\mathrm{glitch}$, to
candidates in the hopes of improving the robustness of existing detection
pipelines. To do this, we first recast the set of SVD SNR timeseries as grayscale
images, shown in Figure~\ref{fig:t_domain}. To help mitigate the high frequency
features, we perform a Fourier Transform to express
the basis responses in the frequency domain:
\begin{equation}
\widehat{Q}_\mu(f_k) = \sum_{j=0}^{N-1} Q_\mu(t_j) e^{-2\pi i f_k t_j}
\end{equation}
where $N$ is the number of data points in our original timeseries, $t_j$
denotes a single time sample, and $f_k$ is a single frequency point. The
positive frequency components are shown in Figure~\ref{fig:f_domain}. The full 
frequency domain images with positive and negative frequency components form the training data we use for the convolutional neural networks (CNNs).

\subsection{Glitch models}
\label{sec:glitchmodels}

There are a wide variety of glitches that impact GW searches, but in this work
we focus on blip glitches~\cite{Cabero:2019orq}, which severely impact compact binary
coalescence searches for sources with total mass $M_\mathrm{tot} \gtrsim 100$,
and scattered light glitches~\cite{LIGO:2020zwl}, which are well approximated by analytic methods
and frequently overlap GW signals. Each of these transients
is well modeled. Although there are several works that have presented more
generic glitch models~\cite{Powell:2022pcg}, they have thus far relied on using images of the
time-frequency morphology to train; the models that we
use are either exact or directly utilize the underlying strain data.

Blip glitches~\cite{Cabero:2019orq} are short
duration $\mathcal{O}(10 \mathrm{ms}$) transients of unknown origin that
appear in both the Advanced LIGO and Advanced Virgo interferometers. They have
a similar morphology to the expected GW signature of certain types of massive
black hole binaries, and they occupy the overlapping frequency band
of $30 - 250$ Hz. 
Some blips exhibit correlations with environmental conditions
or subsystems at the detectors, but the
vast majority lack any known correlation with monitored auxiliary channels.
Although their cause remains unknown, blips are easily identifiable and
cataloging efforts~\cite{Zevin:2016qwy} have confidently identified thousands
of them in LIGO Livingston and LIGO Hanford data. This has facilitated the
development of tools, such as
\texttt{gengli}~\cite{2022PhRvD.106b3027L,2022arXiv220509204L}, that model this
class of glitches.

\texttt{Gengli} uses generative adversarial networks
(GANs)~\cite{2014arXiv1406.2661G} to learn the underlying morphology of blip
glitches, and can accurately
simulate blip glitches that bear both qualitative and statistical similarity
to the underlying training data. The data \texttt{gengli} uses to train is
limited, however, and is heavily dependent on the
GravitySpy~\cite{Zevin:2016qwy} classification scheme. We assume it provides a
faithful representation of blips in the remainder of this work.

While blip glitches have unknown origin, other noise transients are very well
modeled by physical processes within the interferometer.  \emph{Scattered light
glitches} are a consequence of imperfections in the freely falling test masses
at the LIGO detecors. Lasers incident on the mirrors can scatter away from the
main beam path and rejoin after subsequent reflection off other detector
components.  The motion of these detector components relative to the test mass
results in phase shifting of the scattered light, coupling the noise in the
detector to this motion \cite{Accadia:2010zzb, LIGO:2020zwl}.  This results in
long duration transients that occupy the same frequency band as gravitational
wave signals. Although they do not typically mimic GW signals, they were the
most common glitch to overlap with true signals in O3 ($\gtrsim 20\%$ of
confident candidates)~\cite{LIGOScientific:2020ibl,LIGOScientific:2021djp}.  We
consider in particular slow scattering glitches, which result from significant
motion in the microseism band \cite{LIGO:2020zwl, Soni:2024isj}, and produce a
sequence of arches due to multiple bounces from the scatterer.  We approximate
these arches using the model put forth in~\cite{Udall:2022vkv}.
They describe the motion of the test masses as a simple harmonic oscillator,
finding that the induced strain for light scattered $N$ times is:
\begin{equation}
\label{eq:scattering}
h(t) = \sum_{k=0}^N A_k \sin \bigg[ \frac{f_{harm, k}}{f_{mod}} \sin(2\pi f_{mod}(t-t_{c}) + \phi_k) \bigg]
\end{equation}
where $A_k$ denotes the amplitude of the arch, $f_{harm,k}$ is the maximum
frequency of the $k$-th harmonic, $1 / f_{mod}$ is twice the glitch duration, $t_{c}$ is the glitch
end time, and $\phi_k$ is a phase offset for the $k^\mathrm{th}$ arch in the
grouping.

\section{Study}
\label{sec:study}
\subsection{Simulated data}
We generate two BBH template banks with minimum matches~\cite{Owen:1998dk} of
$0.98$ using a stochastic placement
algorithm~\cite{Harry:2009ea,Privitera:2013xza} and a publicly available
estimate of Advanced LIGO's design
sensitivity\footnote{\href{https://dcc.ligo.org/LIGO-T0900288/public}{https://dcc.ligo.org/LIGO-T0900288/public}}.
While constructing the banks, we model the GW emission from 20 Hz using the
phenomenological waveform model \texttt{IMRPhenomD}~\cite{Khan:2015jqa}.  The
first bank is designed to recover
BBHs with morphologies similar to blip glitches. We sample component masses $10
M_\odot < m_{1,2} < 100 M_\odot$ and mandate total masses $60 M_\odot <
M_\mathrm{tot} < 200 M_\odot$. We force the component spins to be anti-aligned
with the orbital angular momentum with magnitude $|s_{1z,2z}|<0.99$. This
results in a bank of 234 waveforms for BBHs with high total mass, asymmetric
masses, and negative $z$-component spins. We will refer to this as the
`glitch-like' bank.

The second bank is designed to be representative of the population of BBHs
Advanced LIGO and Advanced Virgo frequently observe, with component masses $20
M_\odot < m_{1,2} < 40 M_\odot$ and negligible component spin.  This simple criterion results in a bank of
53 waveforms; we will refer to this as the `LVK consistent' bank in this work.
The banks are chosen to cover a limited parameter space to be representative of
how GstLAL segments the space for filtering and background
collection~\cite{Sakon:2022ibh}. Studies with the glitch-like bank will
demonstrate the efficacy of our method in noisy regions of the parameter space,
while the LVK-consistent bank will test generality.

\begin{table}
    \centering
    \caption{\label{table:architecture}CNN Architecture for the glitch (LVK
consistent) BBH banks. We use identical settings for each bank; the networks
differ only due to differences in the input dimension, which is a property of
the decomposed bank.} \begin{tabular}{cccc}
        \hline
        \textbf{Layer Type}  & \textbf{Kernel Size / Units}  & \textbf{Output size}\\ \hline
        Input                & 37(30)x402x1    & -                   \\ \hline
        Conv2D               & 16 filters, 7x7               & 31(24) x 396                    \\ \hline
        MaxPooling           & 2x2                           & 15(12) x 198                    \\ \hline
        Conv2D               & 32 filters, 7x7               & 9(6) x 192                    \\ \hline
        MaxPooling           & 2x2                           & 4(3) x 96                    \\ \hline
        Dense                & 16                            & 1

    \end{tabular}
\end{table}

For each set of templates, we model the emission using the
\texttt{SEOBNRv4\_ROM} waveform approximant~\cite{Bohe:2016gbl}, segment the
waveforms in time~\cite{Cannon:2011vi}, and perform SVD to find a set of
reduced bases~\cite{Cannon:2010qh}. This results in 55 (50) bases distributed
across 2 (2) time slices for the glitch-like (LVK consistent) BBH banks,
respectively. For each bank, the time slice immediately prior to merger
contains the majority of the signal power and is most responsive to deviations
from stationarity around the coalescence. We therefore consider only this
portion of the signal in the rest of this work, which results in 1 time slice
of 37 (30) bases for our banks. 

We likewise consider populations of both glitch-like and LVK consistent
BBHs. For both sets, we generate $10^4$ simulated signals. For the glitch-like
bank, we uniformly draw component masses $m_{1,2} \sim \mathcal{U}(10 M_\odot, 100 M_\odot)$ and force the total
mass
to be $60 M_\odot \leq M_\mathrm{tot} \leq 200 M_\odot$. We allow the $z$-component of
the spin angular momentum to be either aligned or anti-aligned
with the orbital angular momentum, with magnitude $|s_{1z,2z}| \sim \mathcal{U}(0,0.99)$. The
signals are not astrophysically distributed in space; we choose random
locations, orientations, and polarizations for each system, but choose to
distribute the systems uniformly in SNR, $\rho \sim \mathcal{U}(8,100)$, rather than
uniformly in comoving volume. For the LVK consistent BBHs, we use the same SNR
distribution. These simulations, however, are non-spinning and with component masses
uniformly drawn $m_{1,2} \sim \mathcal{U}(20 M_\odot, 40 M_\odot)$.

We also simulate noise transients from the two well-modeled glitch classes
discussed in Section~\ref{sec:glitchmodels}: blip glitches and scattering
arches. We use \texttt{gengli}~\cite{2022arXiv220509204L,2022PhRvD.106b3027L}
to generate $10^4$ blip glitches in the frequency band $f \in [30 \mathrm{Hz},
250 \mathrm{Hz}]$ with SNRs $\rho \sim \mathcal{U}(8,100)$.  For the
scattering arches, we simulate glitches using the model described in Equation~\ref{eq:scattering} from~\cite{Udall:2022vkv},
fixing the number of arches to 3. 
Modulation frequencies are drawn uniformly over the microseism band $f_{mod} \sim \mathcal{U}(0.05 \, \si{Hz}, 0.3 \, \si{Hz})$.
The initial arch has a peak frequency drawn from $f_{harm, 0} \sim \mathcal{U}(16 \, \si{Hz}, 20 \, \si{Hz})$,
and the spacing between subsequent arches is drawn from a distribution $\delta f_{harm} \sim \mathcal{U}(3 \, \si{Hz}, 10 \, \si{Hz})$,
such that the peak frequency of the highest arch varies from $21 \, \si{Hz}$ to $40 \, \si{Hz}$.
We once more draw $10^4$ glitches, and rescale the amplitudes such that
the SNRs are distributed between 8 and 100.

We inject the simulated BBHs and each
glitch class into one month of stationary, Gaussian noise recolored to Advanced
LIGO's design sensitivity; the underlying noise is the same data used
in~\cite{Sachdev:2020lfd}. The simulations are uniformly spaced and do not overlap in time. We also collect the response to Gaussian noise without any injected simulations. We filter each data set using both template banks
described above and record the basis vector responses. We store the orthogonal
SNRs observed by these bases for $0.1 s$ centered around the time of peak
injected strain. Each simulation results in a $ 37 \times 201 \, (30 \times
201)$ matrix for the glitch-like (LVK consistent) bank. The rows, columns, and
values in these matrices represent the basis vector number, time, and
orthogonal SNR measured, respectively. Finally, we recast these matrices as grayscale
images for input into our CNN, and use a Fourier Transform to convert to the frequency domain. We store the real and complex components
in alternating columns, resulting in images with dimension $ 37 \times 402 \, (30 \times 402)$.
In total, we collect five separate sets of output per bank: the responses to
glitch-like BBHs, LVK consistent BBHs, blip glitches, scattering arches, and
Gaussian noise. 

To facilitate comparisons to existing glitch mitigation techniques within
detection pipelines, we additionally record the SNR and $\xi^2$ value
associated with the physical templates at the time of each simulated signal and
glitch. Since there are multiple templates that could identify a given
transient, we prefer the trigger
that maximizes
\begin{equation}
\label{eq:rhobar} 
\bar{\rho} = \frac{\rho}{\left[\frac{1}{2}\left(1 + \max(1, \xi^2)^3\right)\right]^{1/5}}\, , 
\end{equation}
This statistic~\cite{LIGOScientific:2011jth,Babak:2012zx} is presently used in
archival analyses for the same purpose.

\subsection{Binary classifier training and performance}
\label{sec:results}

\begin{figure*}
\includegraphics[width=\textwidth]{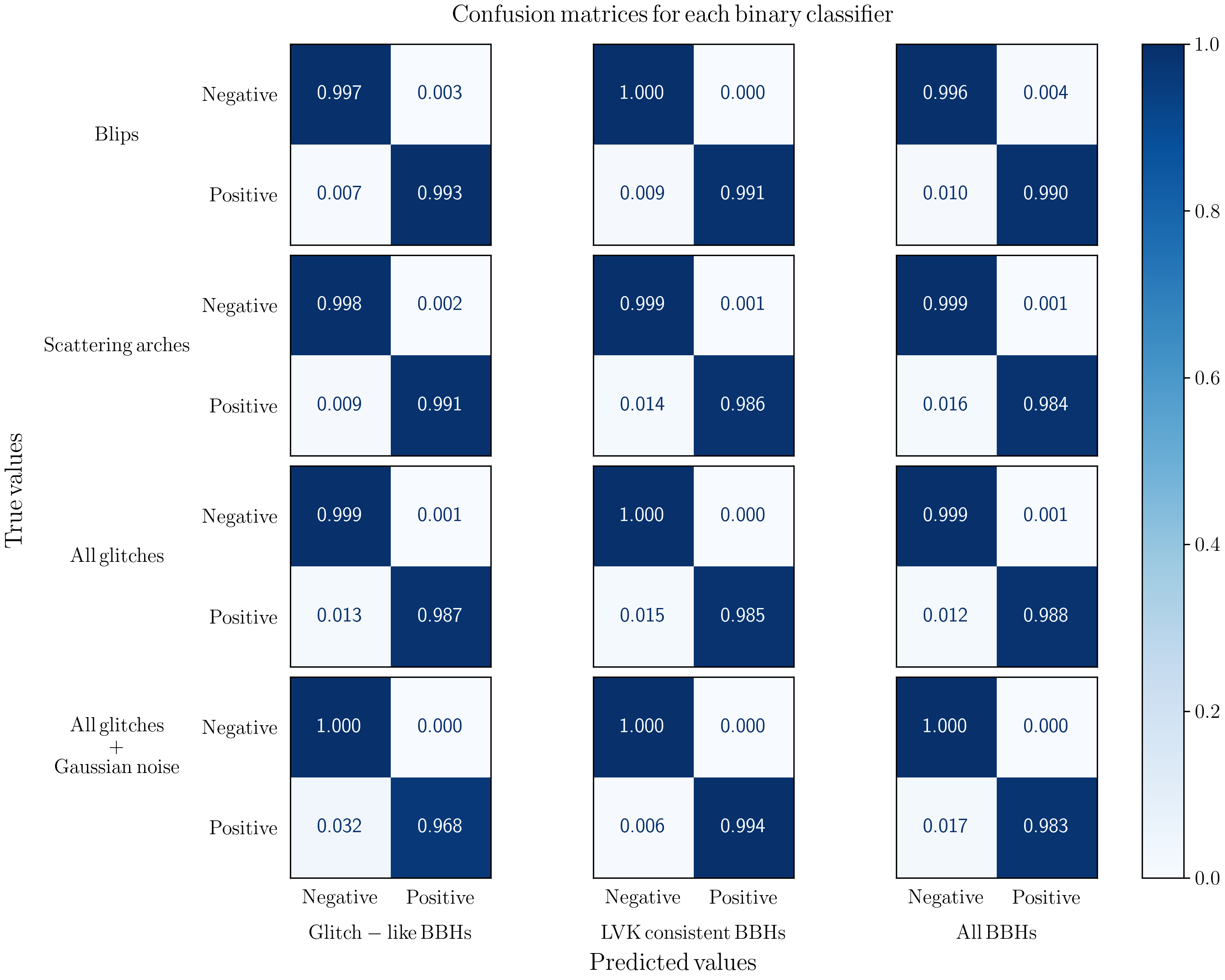}
\caption{\label{fig:confusion_glitch_bank} Confusion matrices for each of the
binary classifiers that we consider for the glitch-like BBH bank. The $x$ and
$y$ axes for each subplot show the false and true values, respectively. The
rows and columns of the entire plot show the two classes compared by the
classifier. The top left corner, for example, shows the performance of the
classifier in distinguishing blip glitches from a glitch-like BBH population.
For this classifier, there is a true positive rate of $99\%$ and a false
positive rate of $0.7\%$.}
\end{figure*}

For both banks, we consider four glitch scenarios and three signal scenarios. In each
case, we ask if a CNN-based classifier can accurately distinguish between the glitch and signal
hypothesis when presented with test data.
The four glitch scenarios include the basis responses to:
\begin{itemize}
\item blip glitches
\item scattering arches
\item blip glitches and scattering arches, combined
\item both glitch classes and Gaussian noise 
\end{itemize}
The three signal scenarios contain the basis response to:
\begin{itemize}
\item high mass, high spin BBHs (glitch-like BBHs)
\item nonspinning $\sim (30 M_\odot, 30 M_\odot)$ BBHs (LVK consistent BBHs)
\item both sets of BBHs combined
\end{itemize}
In total, we construct 12 independent binary classifiers for each bank to
account for every combination of the above data. The input
data, $x_i$, are assigned a classification, $y_i$, of either 0 or 1; here, we
use $1$ to label images containing astrophysical sources and $0$ to label
images that contain purely noise. We pass batches of $N$ input images through a
simple CNN consisting of two convolutional layers, each immediately followed by
a pooling layer, and one fully connected layer.  After each pooling layer, we
use a variant of Rectified Linear Units (ReLUs) known as
Leaky ReLUs, as an activation function:
\begin{equation} f(x) = 
\begin{cases}
x & x > 0 \\
\alpha \cdot x & \mathrm{else}
\end{cases}
\end{equation}
Here, $\alpha$ is a small (non-zero) number. This modification to the standard
ReLU activation function helps to mitigate against vanishing gradients
throughout optimization~\cite{2015arXiv150500853X}. Our output layer 
uses a sigmoid activation function to map our classifier to
the range $[0,1]$, allowing us to interpret the network output,
$y_{\mathrm{pred}, i}$, as a probability.

We separate the data
using a standard 80-10-10 split; 80\% of the data is used to train the network,
10\% is used to validate its performance, and 10\% is reserved for
testing the performance of the final model.
We train using an equal amount of glitch and signal images. Noise
transients, however, are much more frequent than astrophysical signals in
Advanced LIGO data. Blip glitches appeared at a rate of $\sim 50$ per day in
O2~\cite{Cabero:2019orq} and $\sim 100$ per day in O3~\cite{Macas:2022afm}.
By contrast, even in the current observing run, candidate GWs are only identified
approximately once every 2 days\footnote{This is estimated from the number of significant public alerts on \href{https://gracedb.ligo.org/}{\texttt{GraceDb}}.}. To account for the class
imbalance, we compute and minimize a weighted binary cross-entropy (BCE) loss:
\begin{equation}
\begin{split}
\mathrm{BCE \, loss} = -\frac{1}{N} \sum_{i=0}^{N}  & \, w_i \, y_i \, \log y_{\mathrm{pred}, i} \, +\\
& w_i \, (1 - y_i) \, \log(1-y_{\mathrm{pred}, i}) \, .
\end{split}
\end{equation}
using the Adam optimizer~\cite{Kingma:2014vow}. Here, the weight
$w_i$ is a tunable parameter that sets the relative importance of each class.
We apply a weight of $10$ to data samples corresponding to noise transients and
$1$ to those of signals. This allows us to measure the performance of the
network on a distribution akin to the actual data. The specific architecture,
kernel sizes, and output sizes for each layer of our network can be found in
Table~\ref{table:architecture}.

The trained networks output a score that denotes the
probability that a given input corresponds to a signal; we interpret this as
$p_\mathrm{signal}$ and its complement,
\begin{equation}p_\mathrm{glitch} = 1 - p_\mathrm{signal}\end{equation}
as the glitch probability. In order to determine if a given candidate should be
classified as a signal or a glitch, we must choose a decision threshold.
Naively, we might choose to label candidates with $p_\mathrm{signal} \geq
0.5$ as signals, and those with $p_\mathrm{signal} < 0.5$ as glitches. Appropriate decision
thresholds, however, can vary depending on objective and performance desires. Common
considerations include assessing the network's performance on the set of positive predictions. This includes the
\emph{precision}, or fraction of positive predictions that are actually
positive, and the \emph{recall}, or the fraction of positive predictions out of
all positives in the underlying data set. Some statistics, such as the
$\mathrm{F}_1$-score~\cite{van1979information}, incorporate both to approximate
predictive performance.  Other popular measures such as the
informedness~\cite{Youden1950}, Matthews correlation
coefficient~\cite{MATTHEWS1975442}, and Cohen
score~\cite{doi:10.1177/001316446002000104}, additionally incorporate the
performance of negative predictions. 
It is common
to maximize the value of one of these fiducial statistics with respect to the
decision threshold to identify the optimal threshold. In our case, we find that each
classifier produces confident models, with most probabilities close to 0 or 1.
This causes all of the above statistics to remain
approximately constant across decision thresholds. We therefore choose the
naive boundary of 0.5.

\begin{figure*}
\includegraphics[width=\textwidth]{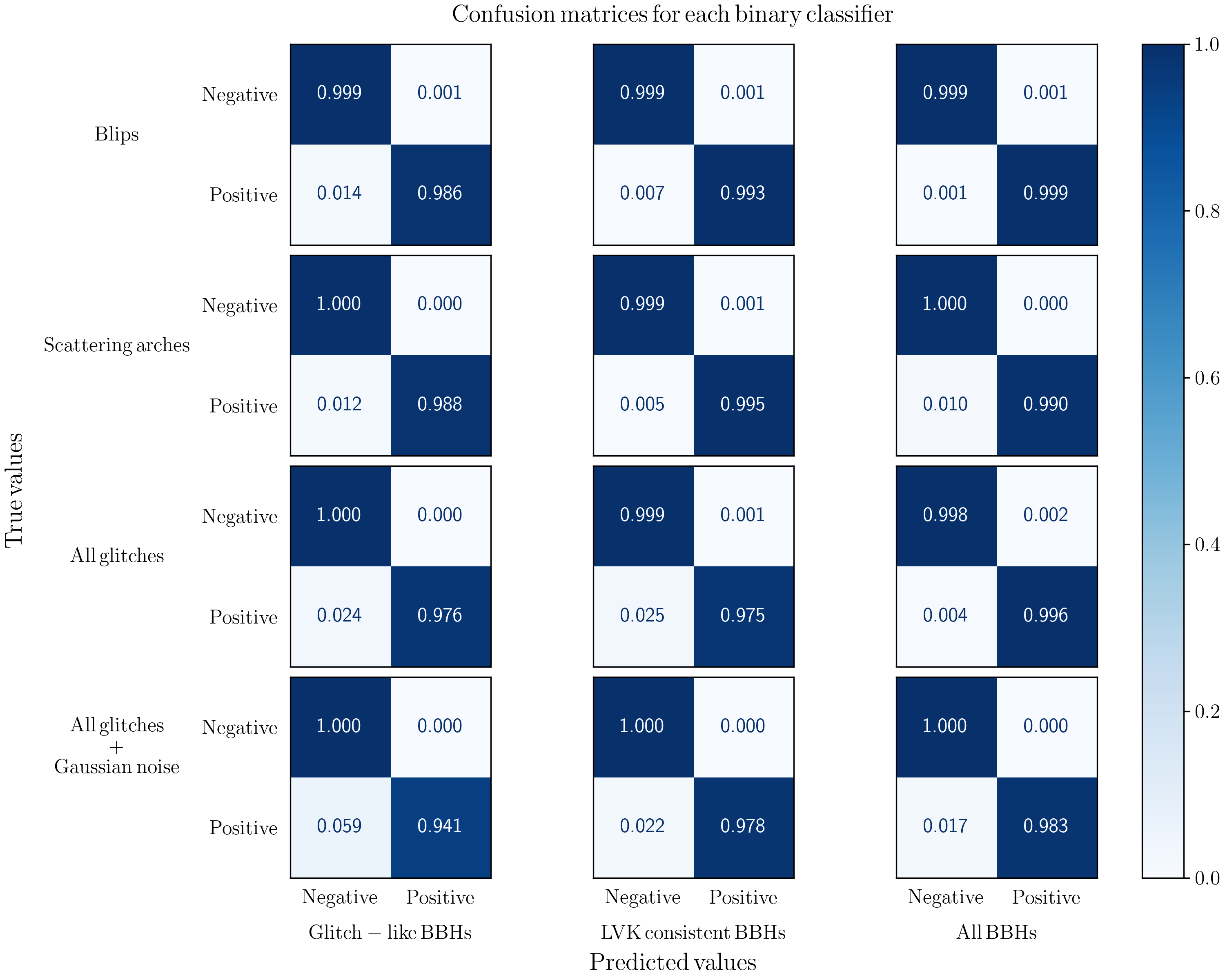}
\caption{\label{fig:confusion_lvk_bank}
Confusion matrices for each of the
binary classifiers that we consider for the LVK consistent BBH bank. The $x$ and
$y$ axes for each subplot show the false and true values, respectively.}
\end{figure*}

Our results using this threshold are summarized by the confusion matrices~\cite{STEHMAN199777} shown in Figures~\ref{fig:confusion_glitch_bank}
and~\ref{fig:confusion_lvk_bank} for the glitch-like and LVK-consistent banks,
respectively.
In general, we find that the binary classifier consistently exhibits a high
true positive rate (TPR) and low false positive rate (FPR).
From the top rows, we see that blip glitches are consistently
separated from GW signals; this is independent of the region of the search parameter space and the nature of the CBCs within the data. The
second row in each figure considers our second fiducial glitch model, scattering arches.
These have a distinct frequency evolution from GW signals (see e.g.
Figures~\ref{fig:t_domain} and~\ref{fig:f_domain}), so we
expect that the response imprinted on the reduced bases will be
distinct as well. Once again, we find that the classifiers perform well for both banks and GW signal classes.
Finally, the last two rows show the performance when both glitch types are
combined with each other and with Gaussian noise. Although this leads to slightly fewer true positives, the network still correctly classifies the vast majority of images. 

We can interpret the first (second) \emph{column} of
Figure~\ref{fig:confusion_glitch_bank} (\ref{fig:confusion_lvk_bank}) as a
measure of the classifier's discerning power when the astrophysical signal
matches the confines of the bank. The second (first) column, on the other hand,
describes the classifier's ability to distinguish BBHs with parameters outside
of its bounds from glitches.  Once again, the CNN performs exceptionally,
allowing us to conclude that CNNs robustly distinguish between the imprint of
glitches and signals on bases that span the CBC space.

\subsection{Improving a toy model detection pipeline}

\begin{figure}
\includegraphics[width=\linewidth]{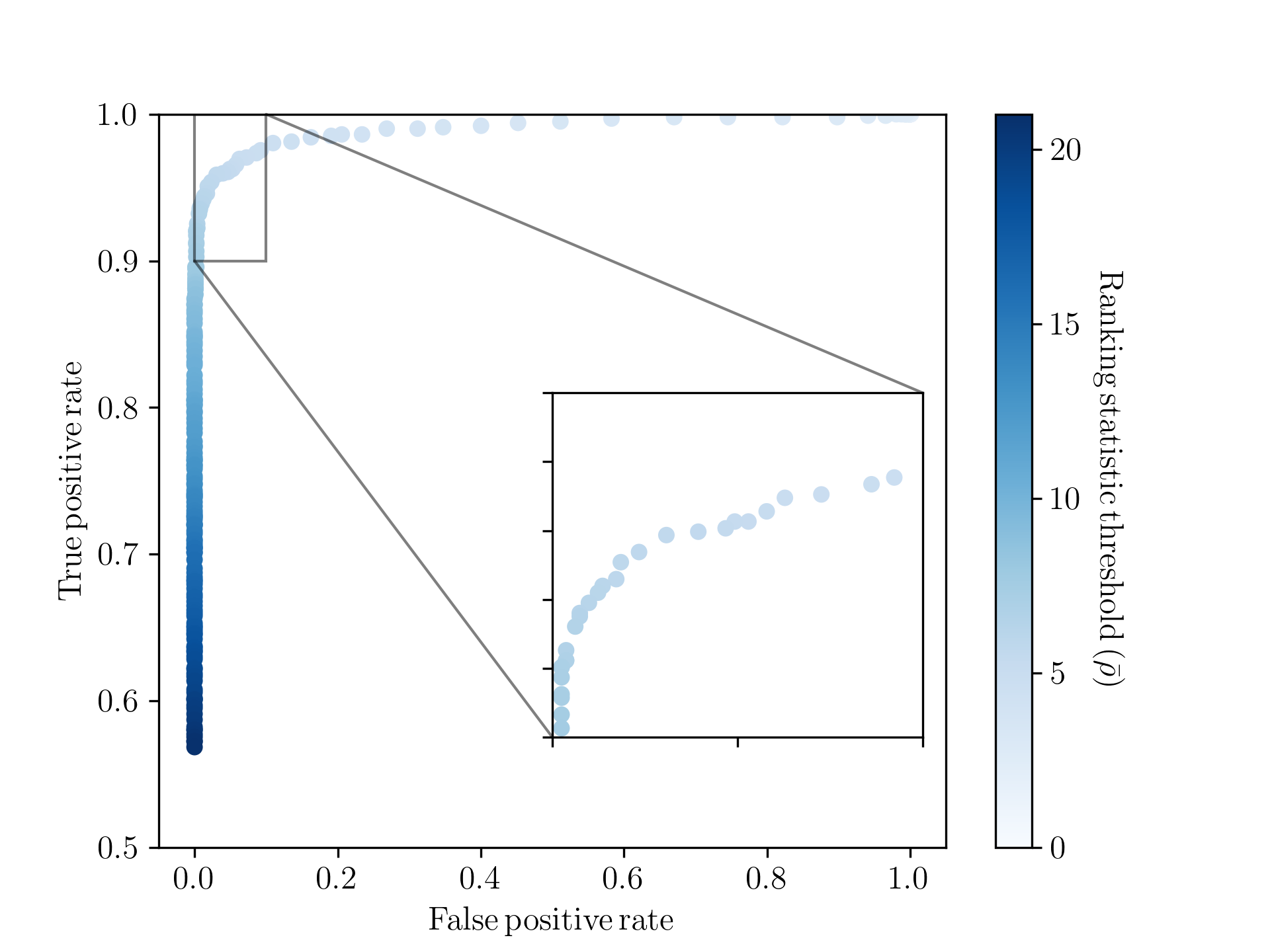}
\caption{\label{fig:toyperf} Receiver operating characteristic (ROC) curve for
our toy model as a function of the decision threshold, $\bar{\rho}$. For
$\bar{\rho} \geq 8$, the toy model produces no false positives in our test set of $\sim 1000$ simulations. The true
positive rate, however, indicates that at this threshold approximately $\sim
10\%$ of astrophysical transients are misclassified.}
\end{figure}

To demonstrate how the $p_\mathrm{glitch}$ produced by our classifier can be
used to improve robustness to noise transients, we construct a simple, toy
model detection pipeline that incorporates both SNR and signal-consistency
information and measure the performance when blip glitches are mixed with
glitch-like astrophysical transients.  Here, we specifically focus on the `glitch-like' bank described in Section~\ref{sec:study}. Following~\cite{Magee:2023muf}, we use
the $\xi^2$-weighted SNR value $\bar{\rho}$, defined in
Equation~\ref{eq:rhobar}, as a ranking statistic. To provide a binary detection
classification, we adopt a selection threshold of $\bar{\rho} \geq
\bar{\rho}_\mathrm{thresh}$.  Signals that exceed this threshold are considered
recovered by the pipeline; those below this threshold have either too low of an
SNR or too poor of a $\xi^2$ consistency check value to be deemed significant.
Explicitly, we define:

\begin{equation}
p_\mathrm{signal}(\bar{\rho}) = \begin{cases} 
      0 & \bar{\rho} < \bar{\rho}_\mathrm{thresh} \\
      1 & \bar{\rho} \geq \bar{\rho}_\mathrm{thresh} 
   \end{cases}
\end{equation}
We measure $\bar{\rho}$ for each
simulated signal and glitch, and impose two distinct detection thresholds
($\bar{\rho} = 7, 10$) to identify marginal and
confident candidates, respectively. Figure~\ref{fig:confusion_pipeline_glitch_bank} shows
the performance of this model.  In general, we find that the toy model does an
excellent job at avoiding false positives, albeit at the cost of producing
fewer true positives. At a modest threshold of $\bar{\rho} = 7$, we find that
$92\%$ of the simulated BBHs are correctly identified, with only $0.25\%$ of all
glitches mistakenly marked as astrophysical in origin. At the more conservative
threshold of $\bar{\rho} = 10$, the true and false positive rates decrease to
$84\%$ and $\lesssim 0.1\%$, respectively. Although the FPR in each case
appears low, we caution that there are $\mathcal{O}(10^3)$ blip glitches per
detector per observing run~\cite{Cabero:2019orq}, and a FPR of
even $0.1\%$ would mistakenly classify several blips per observing run as
astrophysical. The performance of the toy pipeline at a variety of decision
thresholds is shown in Figure~\ref{fig:toyperf}.

Thus far, we demonstrated that training on the reduced basis response produces classifiers that are at least as accurate as this toy pipeline.
As a proof-of-concept demonstration for how the classifier scores might be used mitigate the impact of glitches, we modify our original detection statistic, $\bar{\rho}$, to include the glitch
probability as a weight on the signal consistency check:
\begin{equation}
\label{eq:weighted_lr}
\bar{\rho} = \frac{\rho}{\left[\frac{1}{2}\left(1 + \max(1, \frac{1}{2}(2 \, p_\mathrm{glitch} + 1)\, \xi^2)^3\right)\right]^{1/5}}\, .
\end{equation}
We choose this weighting for a few reasons.
First, as in Equation~\ref{eq:rhobar}, $\bar{\rho} \leq \rho$. 
Second, for $p_\mathrm{glitch} < 0.5$, the network believes the data to be signal-like.
In this case, the signal-consistency check is downweighted so that $\bar{\rho}$
can increase to a maximum value of $\rho$. Alternatively, if $p_\mathrm{glitch}
> 0.5$, the CNN considers the data to be glitch-like and the impact of the signal consistency
check is increased. Finally, if the network is uninformative at that time (e.g.
$p_\mathrm{glitch} = 0.5$), there is no weighting applied and the ranking
statistic is unchanged from the original toy pipeline. This weighting is also resistant to poorly performing classifiers; it can only increase or decrease the assigned $\xi^2$ value by a factor of 2. 

\begin{figure}
\includegraphics[width=\linewidth]{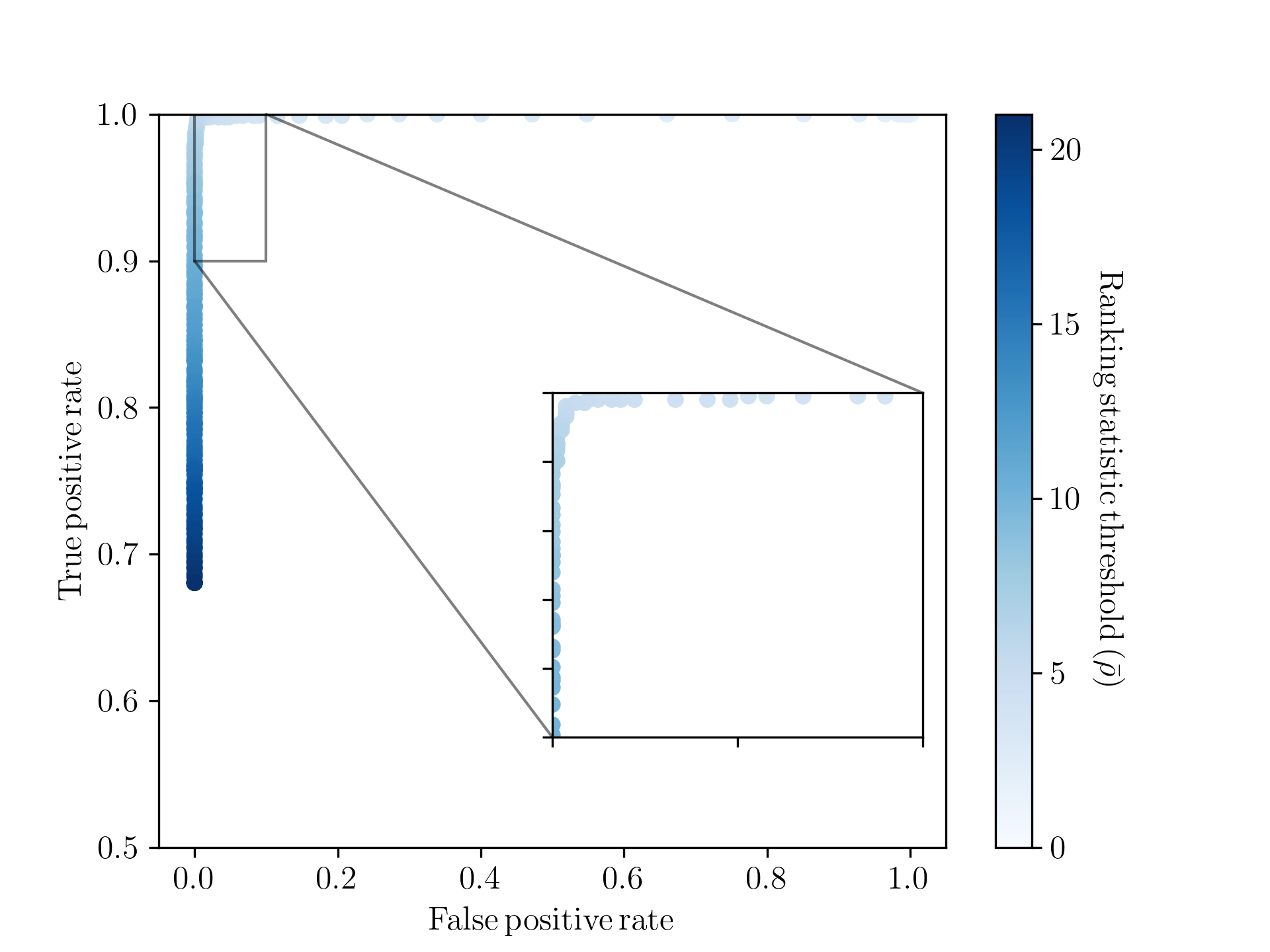}
\caption{\label{fig:toyimprovement} Receiver operating characteristic (ROC) curve for
our $p_\mathrm{glitch}$ weighted toy model as a function of the decision
threshold, $\bar{\rho}$. For $\bar{\rho} \geq 6$, the updated model produces no
false positives. The TPR at this threshold is $100\%$; at
$\bar{\rho} = 8$, this only decreases to $99\%$.}
\end{figure}

We find that applying this weight within the toy pipeline increases the TPR
and the decreases the FPR across detection thresholds. This performance is summarized at a
variety of decision thresholds in Figure~\ref{fig:toyimprovement}. In particular,
we note that at the two thresholds previously considered $(\bar{\rho} = 7, 10$)
the TPR increases from $94\%$ to $98\%$ and $84\%$ to $90\%$, respectively.
The FPR also improves, dropping from $1\%$ to $\lesssim 0.1\%$ for $\bar{\rho}
= 7$ (and remaining below our lower bound of $\lesssim 0.1\%$ for $\bar{\rho} =
10)$.

\begin{figure} 
\centering
\subfloat[]{%
        \includegraphics[width=\linewidth]{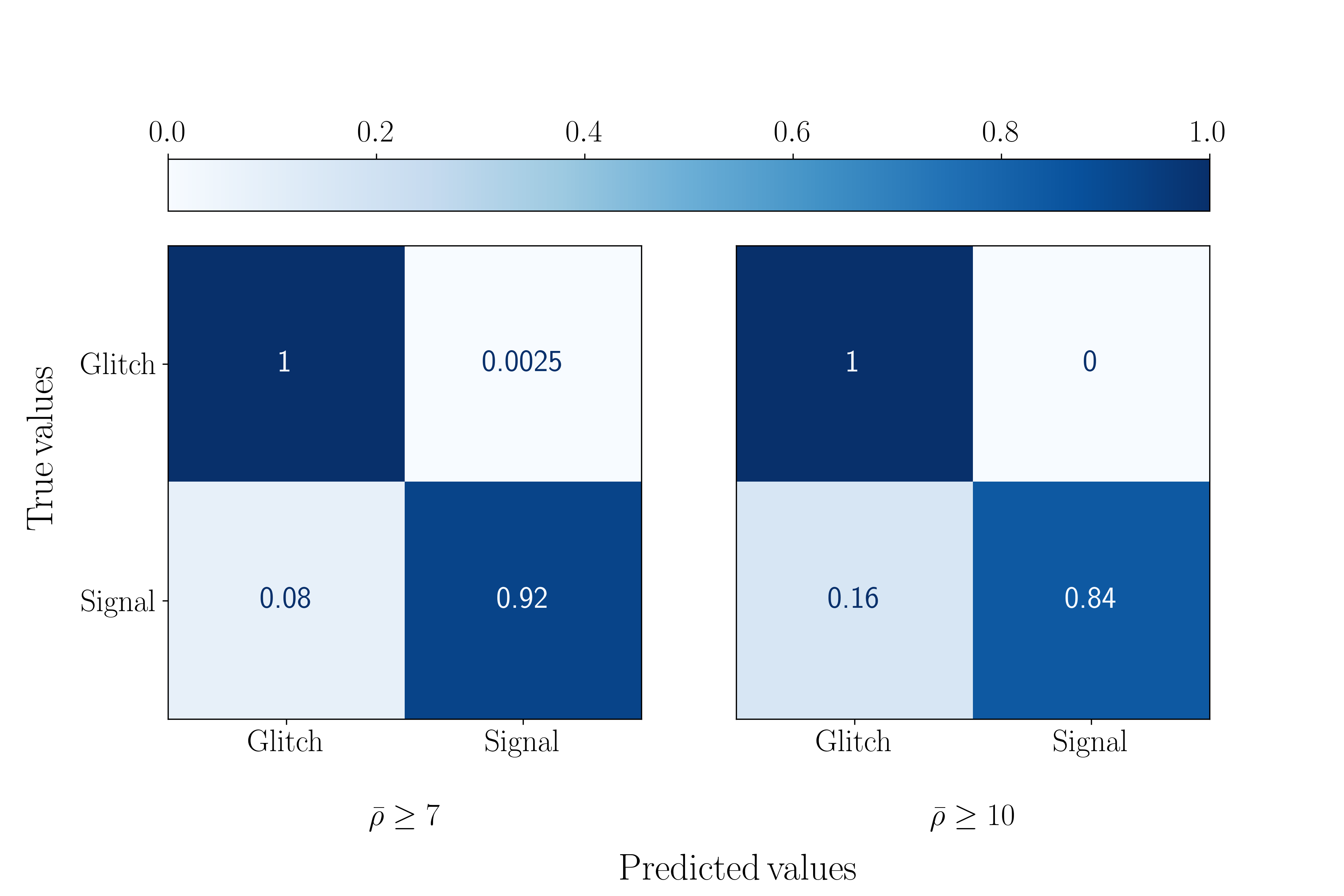}
        \label{fig:confusion_pipeline_glitch_bank}
}
\caption{\label{fig:confusion_pipeline_glitch_bank} 
Confusion matrices for BBH recovery by the toy model pipeline for the
glitch-like bank at 2 fiducial detection statistic thresholds. The false
positive rates are $0.7\%$ and $\lesssim 0.1\%$ our low and high thresholds, respectively. The TPR, however, is more tightly coupled to the detection threshold. While $94\%$ of signals are correctly classified at $\bar{\rho} = 7$, we find that at a
threshold of $\bar{\rho} = 10$ up $16 \%$ of BBHs are missed.}
\end{figure}

\section{Discussion}
\label{sec:conclusions}

Noise transients in Advanced LIGO and Advanced Virgo data can mimic high-mass
BBHs, which remain a target for GW searches despite the expected dearth of
black holes between $\sim 45 M_\odot$ and $\sim 120 M_\odot$ from
pair-instability
supernovae~\cite{Woosley:2016hmi,Belczynski:2016jno,Giacobbo:2017qhh,Spera:2017fyx}.
The detection of a system with component masses in the above range would
bolster existing
observations~\cite{LIGOScientific:2020iuh,LIGOScientific:2020ufj} and help
better constrain binary formation channels. The continued development of noise
mitigation techniques and signal consistency checks for detection pipelines is
therefore crucial to increase the purity of our GW catalogs, and help us to
more robustly identify populations of interest.

In this work, we have demonstrated that a CNN trained to distinguish between
the projection of glitches and signals on a set of reduced bases has a true
positive rate of $\gtrsim 99\%$ across a variety of scenarios. On its own, this
method rejects a similar number of candidates as the toy detection pipeline we
consider. It does, however, reject complementary candidates and exhibit a
higher TPR and lower FPR than the
unaltered pipeline. We show that incorporating the glitch probability estimated
by the classifier into our toy detection pipeline increases its sensitivity
across detection thresholds, suggesting that this method may be worth
integrating more rigorously in production pipelines. 

This study reveals a number of key findings. First, it reinforces that modern
detection pipelines excel at ignoring noise transients. Although blip glitches
and other noise transients can appear significant during single interferometer
times, even the simple detection pipeline we considered rejected the vast
majority of transients.  

Second, we find that CNNs generically separate signals from noise transients.
Since modern search pipelines~\cite{Tsukada:2023edh} partition the GW parameter
space, we considered two distinct regions of the BBH space. Although each
classifier was trained on noise transients and the specific class of BBHs
within the confines of that bank, they still successfully identified GWs from other BBH
regions as signals. This remained true as we expanded the number of transient
classes included in our signal and noise training sets.

Third, when the responses to multiple glitch classes are combined with the
response to Gaussian noise we find that CNNs are still able to identify BBH
signals. This provides early evidence that the basis response alone can be used
for detecting compact binaries, which agrees with previous analytic
work~\cite{Cannon:2011tb}.

Finally, we note that this method is meant to target transients that occur in times with one
operating interferometer, but it is possible to add detectors as an additional
channel in our CNN. We leave studies on the performance of this method in
multiple interferometer time and its coupling to other noise suppression
techniques for future work.

\clearpage

\begin{acknowledgements}
We gratefully acknowledge informative conversations with Cody Messick.
LIGO was constructed by the California Institute of Technology and
Massachusetts Institute of Technology with funding from the National Science
Foundation and operates under cooperative agreement PHY-1764464.
This research has made use of data, software and/or web tools obtained from the
Gravitational Wave Open Science Center (https://www.gw-openscience.org), a
service of LIGO Laboratory, the LIGO Scientific Collaboration and the Virgo
Collaboration.  Virgo is funded by the French Centre National de Recherche
Scientifique (CNRS), the Italian Istituto Nazionale della Fisica Nucleare
(INFN) and the Dutch Nikhef, with contributions by Polish and Hungarian
institutes.
This material is based upon work supported by NSF's LIGO Laboratory which is a
major facility fully funded by the National Science Foundation.
The authors are grateful for computational resources provided by the LIGO
Laboratory and supported by NSF Grants PHY-0757058 and PHY-0823459.
This research has made use of data or software obtained from the Gravitational
Wave Open Science Center (gwosc.org), a service of LIGO Laboratory, the LIGO
Scientific Collaboration, the Virgo Collaboration, and KAGRA. 
This paper carries LIGO document number \dcc{\DocumentID}.
The filtering was performed with the \textsc{GstLAL}
library~\cite{Tsukada:2023edh,2021SoftX..1400680C,Sachdev:2019vvd,Messick:2016aqy},
built on the \textsc{LALSuite} software library~\cite{lalsuite}. 
\end{acknowledgements}
\bibliographystyle{apsrev4-1}
\bibliography{references}

\begin{thebibliography}{108}%
\makeatletter
\providecommand \@ifxundefined [1]{%
 \@ifx{#1\undefined}
}%
\providecommand \@ifnum [1]{%
 \ifnum #1\expandafter \@firstoftwo
 \else \expandafter \@secondoftwo
 \fi
}%
\providecommand \@ifx [1]{%
 \ifx #1\expandafter \@firstoftwo
 \else \expandafter \@secondoftwo
 \fi
}%
\providecommand \natexlab [1]{#1}%
\providecommand \enquote  [1]{``#1''}%
\providecommand \bibnamefont  [1]{#1}%
\providecommand \bibfnamefont [1]{#1}%
\providecommand \citenamefont [1]{#1}%
\providecommand \href@noop [0]{\@secondoftwo}%
\providecommand \href [0]{\begingroup \@sanitize@url \@href}%
\providecommand \@href[1]{\@@startlink{#1}\@@href}%
\providecommand \@@href[1]{\endgroup#1\@@endlink}%
\providecommand \@sanitize@url [0]{\catcode `\\12\catcode `\$12\catcode
  `\&12\catcode `\#12\catcode `\^12\catcode `\_12\catcode `\%12\relax}%
\providecommand \@@startlink[1]{}%
\providecommand \@@endlink[0]{}%
\providecommand \url  [0]{\begingroup\@sanitize@url \@url }%
\providecommand \@url [1]{\endgroup\@href {#1}{\urlprefix }}%
\providecommand \urlprefix  [0]{URL }%
\providecommand \Eprint [0]{\href }%
\providecommand \doibase [0]{http://dx.doi.org/}%
\providecommand \selectlanguage [0]{\@gobble}%
\providecommand \bibinfo  [0]{\@secondoftwo}%
\providecommand \bibfield  [0]{\@secondoftwo}%
\providecommand \translation [1]{[#1]}%
\providecommand \BibitemOpen [0]{}%
\providecommand \bibitemStop [0]{}%
\providecommand \bibitemNoStop [0]{.\EOS\space}%
\providecommand \EOS [0]{\spacefactor3000\relax}%
\providecommand \BibitemShut  [1]{\csname bibitem#1\endcsname}%
\let\auto@bib@innerbib\@empty
\bibitem [{\citenamefont {Aasi}\ \emph {et~al.}(2015)\citenamefont {Aasi} \emph
  {et~al.}}]{LIGOScientific:2014pky}%
  \BibitemOpen
  \bibfield  {author} {\bibinfo {author} {\bibfnamefont {J.}~\bibnamefont
  {Aasi}} \emph {et~al.} (\bibinfo {collaboration} {LIGO Scientific}),\ }\href
  {\doibase 10.1088/0264-9381/32/7/074001} {\bibfield  {journal} {\bibinfo
  {journal} {Class. Quant. Grav.}\ }\textbf {\bibinfo {volume} {32}},\ \bibinfo
  {pages} {074001} (\bibinfo {year} {2015})},\ \Eprint
  {http://arxiv.org/abs/1411.4547} {arXiv:1411.4547 [gr-qc]} \BibitemShut
  {NoStop}%
\bibitem [{\citenamefont {Acernese}\ \emph {et~al.}(2015)\citenamefont
  {Acernese} \emph {et~al.}}]{VIRGO:2014yos}%
  \BibitemOpen
  \bibfield  {author} {\bibinfo {author} {\bibfnamefont {F.}~\bibnamefont
  {Acernese}} \emph {et~al.} (\bibinfo {collaboration} {VIRGO}),\ }\href
  {\doibase 10.1088/0264-9381/32/2/024001} {\bibfield  {journal} {\bibinfo
  {journal} {Class. Quant. Grav.}\ }\textbf {\bibinfo {volume} {32}},\ \bibinfo
  {pages} {024001} (\bibinfo {year} {2015})},\ \Eprint
  {http://arxiv.org/abs/1408.3978} {arXiv:1408.3978 [gr-qc]} \BibitemShut
  {NoStop}%
\bibitem [{\citenamefont {Abbott}\ \emph
  {et~al.}(2021{\natexlab{a}})\citenamefont {Abbott} \emph
  {et~al.}}]{LIGOScientific:2021djp}%
  \BibitemOpen
  \bibfield  {author} {\bibinfo {author} {\bibfnamefont {R.}~\bibnamefont
  {Abbott}} \emph {et~al.} (\bibinfo {collaboration} {LIGO Scientific, VIRGO,
  KAGRA}),\ }\href@noop {} {\  (\bibinfo {year} {2021}{\natexlab{a}})},\
  \Eprint {http://arxiv.org/abs/2111.03606} {arXiv:2111.03606 [gr-qc]}
  \BibitemShut {NoStop}%
\bibitem [{\citenamefont {Abbott}\ \emph {et~al.}(2024)\citenamefont {Abbott}
  \emph {et~al.}}]{LIGOScientific:2021usb}%
  \BibitemOpen
  \bibfield  {author} {\bibinfo {author} {\bibfnamefont {R.}~\bibnamefont
  {Abbott}} \emph {et~al.} (\bibinfo {collaboration} {LIGO Scientific,
  VIRGO}),\ }\href {\doibase 10.1103/PhysRevD.109.022001} {\bibfield  {journal}
  {\bibinfo  {journal} {Phys. Rev. D}\ }\textbf {\bibinfo {volume} {109}},\
  \bibinfo {pages} {022001} (\bibinfo {year} {2024})},\ \Eprint
  {http://arxiv.org/abs/2108.01045} {arXiv:2108.01045 [gr-qc]} \BibitemShut
  {NoStop}%
\bibitem [{\citenamefont {Abbott}\ \emph
  {et~al.}(2021{\natexlab{b}})\citenamefont {Abbott} \emph
  {et~al.}}]{LIGOScientific:2020ibl}%
  \BibitemOpen
  \bibfield  {author} {\bibinfo {author} {\bibfnamefont {R.}~\bibnamefont
  {Abbott}} \emph {et~al.} (\bibinfo {collaboration} {LIGO Scientific,
  Virgo}),\ }\href {\doibase 10.1103/PhysRevX.11.021053} {\bibfield  {journal}
  {\bibinfo  {journal} {Phys. Rev. X}\ }\textbf {\bibinfo {volume} {11}},\
  \bibinfo {pages} {021053} (\bibinfo {year} {2021}{\natexlab{b}})},\ \Eprint
  {http://arxiv.org/abs/2010.14527} {arXiv:2010.14527 [gr-qc]} \BibitemShut
  {NoStop}%
\bibitem [{\citenamefont {Abbott}\ \emph
  {et~al.}(2019{\natexlab{a}})\citenamefont {Abbott} \emph
  {et~al.}}]{LIGOScientific:2018mvr}%
  \BibitemOpen
  \bibfield  {author} {\bibinfo {author} {\bibfnamefont {B.~P.}\ \bibnamefont
  {Abbott}} \emph {et~al.} (\bibinfo {collaboration} {LIGO Scientific,
  Virgo}),\ }\href {\doibase 10.1103/PhysRevX.9.031040} {\bibfield  {journal}
  {\bibinfo  {journal} {Phys. Rev. X}\ }\textbf {\bibinfo {volume} {9}},\
  \bibinfo {pages} {031040} (\bibinfo {year} {2019}{\natexlab{a}})},\ \Eprint
  {http://arxiv.org/abs/1811.12907} {arXiv:1811.12907 [astro-ph.HE]}
  \BibitemShut {NoStop}%
\bibitem [{\citenamefont {Nitz}\ \emph {et~al.}(2019)\citenamefont {Nitz},
  \citenamefont {Capano}, \citenamefont {Nielsen}, \citenamefont {Reyes},
  \citenamefont {White}, \citenamefont {Brown},\ and\ \citenamefont
  {Krishnan}}]{Nitz:2018imz}%
  \BibitemOpen
  \bibfield  {author} {\bibinfo {author} {\bibfnamefont {A.~H.}\ \bibnamefont
  {Nitz}}, \bibinfo {author} {\bibfnamefont {C.}~\bibnamefont {Capano}},
  \bibinfo {author} {\bibfnamefont {A.~B.}\ \bibnamefont {Nielsen}}, \bibinfo
  {author} {\bibfnamefont {S.}~\bibnamefont {Reyes}}, \bibinfo {author}
  {\bibfnamefont {R.}~\bibnamefont {White}}, \bibinfo {author} {\bibfnamefont
  {D.~A.}\ \bibnamefont {Brown}}, \ and\ \bibinfo {author} {\bibfnamefont
  {B.}~\bibnamefont {Krishnan}},\ }\href {\doibase 10.3847/1538-4357/ab0108}
  {\bibfield  {journal} {\bibinfo  {journal} {Astrophys. J.}\ }\textbf
  {\bibinfo {volume} {872}},\ \bibinfo {pages} {195} (\bibinfo {year}
  {2019})},\ \Eprint {http://arxiv.org/abs/1811.01921} {arXiv:1811.01921
  [gr-qc]} \BibitemShut {NoStop}%
\bibitem [{\citenamefont {Magee}\ \emph {et~al.}(2019)\citenamefont {Magee}
  \emph {et~al.}}]{Magee:2019vmb}%
  \BibitemOpen
  \bibfield  {author} {\bibinfo {author} {\bibfnamefont {R.}~\bibnamefont
  {Magee}} \emph {et~al.},\ }\href {\doibase 10.3847/2041-8213/ab20cf}
  {\bibfield  {journal} {\bibinfo  {journal} {Astrophys. J. Lett.}\ }\textbf
  {\bibinfo {volume} {878}},\ \bibinfo {pages} {L17} (\bibinfo {year}
  {2019})},\ \Eprint {http://arxiv.org/abs/1901.09884} {arXiv:1901.09884
  [gr-qc]} \BibitemShut {NoStop}%
\bibitem [{\citenamefont {Venumadhav}\ \emph {et~al.}(2020)\citenamefont
  {Venumadhav}, \citenamefont {Zackay}, \citenamefont {Roulet}, \citenamefont
  {Dai},\ and\ \citenamefont {Zaldarriaga}}]{Venumadhav:2019lyq}%
  \BibitemOpen
  \bibfield  {author} {\bibinfo {author} {\bibfnamefont {T.}~\bibnamefont
  {Venumadhav}}, \bibinfo {author} {\bibfnamefont {B.}~\bibnamefont {Zackay}},
  \bibinfo {author} {\bibfnamefont {J.}~\bibnamefont {Roulet}}, \bibinfo
  {author} {\bibfnamefont {L.}~\bibnamefont {Dai}}, \ and\ \bibinfo {author}
  {\bibfnamefont {M.}~\bibnamefont {Zaldarriaga}},\ }\href {\doibase
  10.1103/PhysRevD.101.083030} {\bibfield  {journal} {\bibinfo  {journal}
  {Phys. Rev. D}\ }\textbf {\bibinfo {volume} {101}},\ \bibinfo {pages}
  {083030} (\bibinfo {year} {2020})},\ \Eprint
  {http://arxiv.org/abs/1904.07214} {arXiv:1904.07214 [astro-ph.HE]}
  \BibitemShut {NoStop}%
\bibitem [{\citenamefont {Zackay}\ \emph {et~al.}(2021)\citenamefont {Zackay},
  \citenamefont {Dai}, \citenamefont {Venumadhav}, \citenamefont {Roulet},\
  and\ \citenamefont {Zaldarriaga}}]{Zackay:2019btq}%
  \BibitemOpen
  \bibfield  {author} {\bibinfo {author} {\bibfnamefont {B.}~\bibnamefont
  {Zackay}}, \bibinfo {author} {\bibfnamefont {L.}~\bibnamefont {Dai}},
  \bibinfo {author} {\bibfnamefont {T.}~\bibnamefont {Venumadhav}}, \bibinfo
  {author} {\bibfnamefont {J.}~\bibnamefont {Roulet}}, \ and\ \bibinfo {author}
  {\bibfnamefont {M.}~\bibnamefont {Zaldarriaga}},\ }\href {\doibase
  10.1103/PhysRevD.104.063030} {\bibfield  {journal} {\bibinfo  {journal}
  {Phys. Rev. D}\ }\textbf {\bibinfo {volume} {104}},\ \bibinfo {pages}
  {063030} (\bibinfo {year} {2021})},\ \Eprint
  {http://arxiv.org/abs/1910.09528} {arXiv:1910.09528 [astro-ph.HE]}
  \BibitemShut {NoStop}%
\bibitem [{\citenamefont {Nitz}\ \emph {et~al.}(2020)\citenamefont {Nitz},
  \citenamefont {Dent}, \citenamefont {Davies}, \citenamefont {Kumar},
  \citenamefont {Capano}, \citenamefont {Harry}, \citenamefont {Mozzon},
  \citenamefont {Nuttall}, \citenamefont {Lundgren},\ and\ \citenamefont
  {T\'apai}}]{Nitz:2020oeq}%
  \BibitemOpen
  \bibfield  {author} {\bibinfo {author} {\bibfnamefont {A.~H.}\ \bibnamefont
  {Nitz}}, \bibinfo {author} {\bibfnamefont {T.}~\bibnamefont {Dent}}, \bibinfo
  {author} {\bibfnamefont {G.~S.}\ \bibnamefont {Davies}}, \bibinfo {author}
  {\bibfnamefont {S.}~\bibnamefont {Kumar}}, \bibinfo {author} {\bibfnamefont
  {C.~D.}\ \bibnamefont {Capano}}, \bibinfo {author} {\bibfnamefont
  {I.}~\bibnamefont {Harry}}, \bibinfo {author} {\bibfnamefont
  {S.}~\bibnamefont {Mozzon}}, \bibinfo {author} {\bibfnamefont
  {L.}~\bibnamefont {Nuttall}}, \bibinfo {author} {\bibfnamefont
  {A.}~\bibnamefont {Lundgren}}, \ and\ \bibinfo {author} {\bibfnamefont
  {M.}~\bibnamefont {T\'apai}},\ }\href {\doibase 10.3847/1538-4357/ab733f}
  {\bibfield  {journal} {\bibinfo  {journal} {Astrophys. J.}\ }\textbf
  {\bibinfo {volume} {891}},\ \bibinfo {pages} {123} (\bibinfo {year}
  {2020})},\ \Eprint {http://arxiv.org/abs/1910.05331} {arXiv:1910.05331
  [astro-ph.HE]} \BibitemShut {NoStop}%
\bibitem [{\citenamefont {Nitz}\ \emph {et~al.}(2021)\citenamefont {Nitz},
  \citenamefont {Capano}, \citenamefont {Kumar}, \citenamefont {Wang},
  \citenamefont {Kastha}, \citenamefont {Sch\"afer}, \citenamefont
  {Dhurkunde},\ and\ \citenamefont {Cabero}}]{Nitz:2021uxj}%
  \BibitemOpen
  \bibfield  {author} {\bibinfo {author} {\bibfnamefont {A.~H.}\ \bibnamefont
  {Nitz}}, \bibinfo {author} {\bibfnamefont {C.~D.}\ \bibnamefont {Capano}},
  \bibinfo {author} {\bibfnamefont {S.}~\bibnamefont {Kumar}}, \bibinfo
  {author} {\bibfnamefont {Y.-F.}\ \bibnamefont {Wang}}, \bibinfo {author}
  {\bibfnamefont {S.}~\bibnamefont {Kastha}}, \bibinfo {author} {\bibfnamefont
  {M.}~\bibnamefont {Sch\"afer}}, \bibinfo {author} {\bibfnamefont
  {R.}~\bibnamefont {Dhurkunde}}, \ and\ \bibinfo {author} {\bibfnamefont
  {M.}~\bibnamefont {Cabero}},\ }\href {\doibase 10.3847/1538-4357/ac1c03}
  {\bibfield  {journal} {\bibinfo  {journal} {Astrophys. J.}\ }\textbf
  {\bibinfo {volume} {922}},\ \bibinfo {pages} {76} (\bibinfo {year} {2021})},\
  \Eprint {http://arxiv.org/abs/2105.09151} {arXiv:2105.09151 [astro-ph.HE]}
  \BibitemShut {NoStop}%
\bibitem [{\citenamefont {Nitz}\ \emph {et~al.}(2023)\citenamefont {Nitz},
  \citenamefont {Kumar}, \citenamefont {Wang}, \citenamefont {Kastha},
  \citenamefont {Wu}, \citenamefont {Sch\"afer}, \citenamefont {Dhurkunde},\
  and\ \citenamefont {Capano}}]{Nitz:2021zwj}%
  \BibitemOpen
  \bibfield  {author} {\bibinfo {author} {\bibfnamefont {A.~H.}\ \bibnamefont
  {Nitz}}, \bibinfo {author} {\bibfnamefont {S.}~\bibnamefont {Kumar}},
  \bibinfo {author} {\bibfnamefont {Y.-F.}\ \bibnamefont {Wang}}, \bibinfo
  {author} {\bibfnamefont {S.}~\bibnamefont {Kastha}}, \bibinfo {author}
  {\bibfnamefont {S.}~\bibnamefont {Wu}}, \bibinfo {author} {\bibfnamefont
  {M.}~\bibnamefont {Sch\"afer}}, \bibinfo {author} {\bibfnamefont
  {R.}~\bibnamefont {Dhurkunde}}, \ and\ \bibinfo {author} {\bibfnamefont
  {C.~D.}\ \bibnamefont {Capano}},\ }\href {\doibase 10.3847/1538-4357/aca591}
  {\bibfield  {journal} {\bibinfo  {journal} {Astrophys. J.}\ }\textbf
  {\bibinfo {volume} {946}},\ \bibinfo {pages} {59} (\bibinfo {year} {2023})},\
  \Eprint {http://arxiv.org/abs/2112.06878} {arXiv:2112.06878 [astro-ph.HE]}
  \BibitemShut {NoStop}%
\bibitem [{\citenamefont {Olsen}\ \emph {et~al.}(2022)\citenamefont {Olsen},
  \citenamefont {Venumadhav}, \citenamefont {Mushkin}, \citenamefont {Roulet},
  \citenamefont {Zackay},\ and\ \citenamefont {Zaldarriaga}}]{Olsen:2022pin}%
  \BibitemOpen
  \bibfield  {author} {\bibinfo {author} {\bibfnamefont {S.}~\bibnamefont
  {Olsen}}, \bibinfo {author} {\bibfnamefont {T.}~\bibnamefont {Venumadhav}},
  \bibinfo {author} {\bibfnamefont {J.}~\bibnamefont {Mushkin}}, \bibinfo
  {author} {\bibfnamefont {J.}~\bibnamefont {Roulet}}, \bibinfo {author}
  {\bibfnamefont {B.}~\bibnamefont {Zackay}}, \ and\ \bibinfo {author}
  {\bibfnamefont {M.}~\bibnamefont {Zaldarriaga}},\ }\href {\doibase
  10.1103/PhysRevD.106.043009} {\bibfield  {journal} {\bibinfo  {journal}
  {Phys. Rev. D}\ }\textbf {\bibinfo {volume} {106}},\ \bibinfo {pages}
  {043009} (\bibinfo {year} {2022})},\ \Eprint
  {http://arxiv.org/abs/2201.02252} {arXiv:2201.02252 [astro-ph.HE]}
  \BibitemShut {NoStop}%
\bibitem [{\citenamefont {Abbott}\ \emph {et~al.}(2023)\citenamefont {Abbott}
  \emph {et~al.}}]{KAGRA:2021duu}%
  \BibitemOpen
  \bibfield  {author} {\bibinfo {author} {\bibfnamefont {R.}~\bibnamefont
  {Abbott}} \emph {et~al.} (\bibinfo {collaboration} {KAGRA, VIRGO, LIGO
  Scientific}),\ }\href {\doibase 10.1103/PhysRevX.13.011048} {\bibfield
  {journal} {\bibinfo  {journal} {Phys. Rev. X}\ }\textbf {\bibinfo {volume}
  {13}},\ \bibinfo {pages} {011048} (\bibinfo {year} {2023})},\ \Eprint
  {http://arxiv.org/abs/2111.03634} {arXiv:2111.03634 [astro-ph.HE]}
  \BibitemShut {NoStop}%
\bibitem [{\citenamefont {Abbott}\ \emph
  {et~al.}(2019{\natexlab{b}})\citenamefont {Abbott} \emph
  {et~al.}}]{LIGOScientific:2018dkp}%
  \BibitemOpen
  \bibfield  {author} {\bibinfo {author} {\bibfnamefont {B.~P.}\ \bibnamefont
  {Abbott}} \emph {et~al.} (\bibinfo {collaboration} {LIGO Scientific,
  Virgo}),\ }\href {\doibase 10.1103/PhysRevLett.123.011102} {\bibfield
  {journal} {\bibinfo  {journal} {Phys. Rev. Lett.}\ }\textbf {\bibinfo
  {volume} {123}},\ \bibinfo {pages} {011102} (\bibinfo {year}
  {2019}{\natexlab{b}})},\ \Eprint {http://arxiv.org/abs/1811.00364}
  {arXiv:1811.00364 [gr-qc]} \BibitemShut {NoStop}%
\bibitem [{\citenamefont {Abbott}\ \emph
  {et~al.}(2021{\natexlab{c}})\citenamefont {Abbott} \emph
  {et~al.}}]{LIGOScientific:2021sio}%
  \BibitemOpen
  \bibfield  {author} {\bibinfo {author} {\bibfnamefont {R.}~\bibnamefont
  {Abbott}} \emph {et~al.} (\bibinfo {collaboration} {LIGO Scientific, VIRGO,
  KAGRA}),\ }\href@noop {} {\  (\bibinfo {year} {2021}{\natexlab{c}})},\
  \Eprint {http://arxiv.org/abs/2112.06861} {arXiv:2112.06861 [gr-qc]}
  \BibitemShut {NoStop}%
\bibitem [{\citenamefont {Abbott}\ \emph
  {et~al.}(2018{\natexlab{a}})\citenamefont {Abbott} \emph
  {et~al.}}]{LIGOScientific:2018cki}%
  \BibitemOpen
  \bibfield  {author} {\bibinfo {author} {\bibfnamefont {B.~P.}\ \bibnamefont
  {Abbott}} \emph {et~al.} (\bibinfo {collaboration} {LIGO Scientific,
  Virgo}),\ }\href {\doibase 10.1103/PhysRevLett.121.161101} {\bibfield
  {journal} {\bibinfo  {journal} {Phys. Rev. Lett.}\ }\textbf {\bibinfo
  {volume} {121}},\ \bibinfo {pages} {161101} (\bibinfo {year}
  {2018}{\natexlab{a}})},\ \Eprint {http://arxiv.org/abs/1805.11581}
  {arXiv:1805.11581 [gr-qc]} \BibitemShut {NoStop}%
\bibitem [{\citenamefont {Abbott}\ \emph
  {et~al.}(2020{\natexlab{a}})\citenamefont {Abbott} \emph
  {et~al.}}]{LIGOScientific:2020aai}%
  \BibitemOpen
  \bibfield  {author} {\bibinfo {author} {\bibfnamefont {B.~P.}\ \bibnamefont
  {Abbott}} \emph {et~al.} (\bibinfo {collaboration} {LIGO Scientific,
  Virgo}),\ }\href {\doibase 10.3847/2041-8213/ab75f5} {\bibfield  {journal}
  {\bibinfo  {journal} {Astrophys. J. Lett.}\ }\textbf {\bibinfo {volume}
  {892}},\ \bibinfo {pages} {L3} (\bibinfo {year} {2020}{\natexlab{a}})},\
  \Eprint {http://arxiv.org/abs/2001.01761} {arXiv:2001.01761 [astro-ph.HE]}
  \BibitemShut {NoStop}%
\bibitem [{\citenamefont {Abbott}\ \emph
  {et~al.}(2020{\natexlab{b}})\citenamefont {Abbott} \emph
  {et~al.}}]{LIGOScientific:2020zkf}%
  \BibitemOpen
  \bibfield  {author} {\bibinfo {author} {\bibfnamefont {R.}~\bibnamefont
  {Abbott}} \emph {et~al.} (\bibinfo {collaboration} {LIGO Scientific,
  Virgo}),\ }\href {\doibase 10.3847/2041-8213/ab960f} {\bibfield  {journal}
  {\bibinfo  {journal} {Astrophys. J. Lett.}\ }\textbf {\bibinfo {volume}
  {896}},\ \bibinfo {pages} {L44} (\bibinfo {year} {2020}{\natexlab{b}})},\
  \Eprint {http://arxiv.org/abs/2006.12611} {arXiv:2006.12611 [astro-ph.HE]}
  \BibitemShut {NoStop}%
\bibitem [{\citenamefont {Abbott}\ \emph
  {et~al.}(2020{\natexlab{c}})\citenamefont {Abbott} \emph
  {et~al.}}]{LIGOScientific:2020iuh}%
  \BibitemOpen
  \bibfield  {author} {\bibinfo {author} {\bibfnamefont {R.}~\bibnamefont
  {Abbott}} \emph {et~al.} (\bibinfo {collaboration} {LIGO Scientific,
  Virgo}),\ }\href {\doibase 10.1103/PhysRevLett.125.101102} {\bibfield
  {journal} {\bibinfo  {journal} {Phys. Rev. Lett.}\ }\textbf {\bibinfo
  {volume} {125}},\ \bibinfo {pages} {101102} (\bibinfo {year}
  {2020}{\natexlab{c}})},\ \Eprint {http://arxiv.org/abs/2009.01075}
  {arXiv:2009.01075 [gr-qc]} \BibitemShut {NoStop}%
\bibitem [{\citenamefont {Abac}\ \emph {et~al.}(2024)\citenamefont {Abac} \emph
  {et~al.}}]{LIGOScientific:2024elc}%
  \BibitemOpen
  \bibfield  {author} {\bibinfo {author} {\bibfnamefont {A.~G.}\ \bibnamefont
  {Abac}} \emph {et~al.} (\bibinfo {collaboration} {LIGO Scientific, VIRGO,
  KAGRA}),\ }\href@noop {} {\  (\bibinfo {year} {2024})},\ \Eprint
  {http://arxiv.org/abs/2404.04248} {arXiv:2404.04248 [astro-ph.HE]}
  \BibitemShut {NoStop}%
\bibitem [{\citenamefont {Zevin}\ \emph {et~al.}(2021)\citenamefont {Zevin},
  \citenamefont {Bavera}, \citenamefont {Berry}, \citenamefont {Kalogera},
  \citenamefont {Fragos}, \citenamefont {Marchant}, \citenamefont {Rodriguez},
  \citenamefont {Antonini}, \citenamefont {Holz},\ and\ \citenamefont
  {Pankow}}]{Zevin:2020gbd}%
  \BibitemOpen
  \bibfield  {author} {\bibinfo {author} {\bibfnamefont {M.}~\bibnamefont
  {Zevin}}, \bibinfo {author} {\bibfnamefont {S.~S.}\ \bibnamefont {Bavera}},
  \bibinfo {author} {\bibfnamefont {C.~P.~L.}\ \bibnamefont {Berry}}, \bibinfo
  {author} {\bibfnamefont {V.}~\bibnamefont {Kalogera}}, \bibinfo {author}
  {\bibfnamefont {T.}~\bibnamefont {Fragos}}, \bibinfo {author} {\bibfnamefont
  {P.}~\bibnamefont {Marchant}}, \bibinfo {author} {\bibfnamefont {C.~L.}\
  \bibnamefont {Rodriguez}}, \bibinfo {author} {\bibfnamefont {F.}~\bibnamefont
  {Antonini}}, \bibinfo {author} {\bibfnamefont {D.~E.}\ \bibnamefont {Holz}},
  \ and\ \bibinfo {author} {\bibfnamefont {C.}~\bibnamefont {Pankow}},\ }\href
  {\doibase 10.3847/1538-4357/abe40e} {\bibfield  {journal} {\bibinfo
  {journal} {Astrophys. J.}\ }\textbf {\bibinfo {volume} {910}},\ \bibinfo
  {pages} {152} (\bibinfo {year} {2021})},\ \Eprint
  {http://arxiv.org/abs/2011.10057} {arXiv:2011.10057 [astro-ph.HE]}
  \BibitemShut {NoStop}%
\bibitem [{\citenamefont {Messenger}\ and\ \citenamefont
  {Veitch}(2013)}]{Messenger:2012jy}%
  \BibitemOpen
  \bibfield  {author} {\bibinfo {author} {\bibfnamefont {C.}~\bibnamefont
  {Messenger}}\ and\ \bibinfo {author} {\bibfnamefont {J.}~\bibnamefont
  {Veitch}},\ }\href {\doibase 10.1088/1367-2630/15/5/053027} {\bibfield
  {journal} {\bibinfo  {journal} {New J. Phys.}\ }\textbf {\bibinfo {volume}
  {15}},\ \bibinfo {pages} {053027} (\bibinfo {year} {2013})},\ \Eprint
  {http://arxiv.org/abs/1206.3461} {arXiv:1206.3461 [astro-ph.IM]} \BibitemShut
  {NoStop}%
\bibitem [{\citenamefont {Magee}\ \emph
  {et~al.}(2024{\natexlab{a}})\citenamefont {Magee}, \citenamefont {Isi},
  \citenamefont {Payne}, \citenamefont {Chatziioannou}, \citenamefont {Farr},
  \citenamefont {Pratten},\ and\ \citenamefont {Vitale}}]{Magee:2023muf}%
  \BibitemOpen
  \bibfield  {author} {\bibinfo {author} {\bibfnamefont {R.}~\bibnamefont
  {Magee}}, \bibinfo {author} {\bibfnamefont {M.}~\bibnamefont {Isi}}, \bibinfo
  {author} {\bibfnamefont {E.}~\bibnamefont {Payne}}, \bibinfo {author}
  {\bibfnamefont {K.}~\bibnamefont {Chatziioannou}}, \bibinfo {author}
  {\bibfnamefont {W.~M.}\ \bibnamefont {Farr}}, \bibinfo {author}
  {\bibfnamefont {G.}~\bibnamefont {Pratten}}, \ and\ \bibinfo {author}
  {\bibfnamefont {S.}~\bibnamefont {Vitale}},\ }\href {\doibase
  10.1103/PhysRevD.109.023014} {\bibfield  {journal} {\bibinfo  {journal}
  {Phys. Rev. D}\ }\textbf {\bibinfo {volume} {109}},\ \bibinfo {pages}
  {023014} (\bibinfo {year} {2024}{\natexlab{a}})},\ \Eprint
  {http://arxiv.org/abs/2311.03656} {arXiv:2311.03656 [gr-qc]} \BibitemShut
  {NoStop}%
\bibitem [{\citenamefont {Abbott}\ \emph
  {et~al.}(2018{\natexlab{b}})\citenamefont {Abbott} \emph
  {et~al.}}]{LIGOScientific:2017tza}%
  \BibitemOpen
  \bibfield  {author} {\bibinfo {author} {\bibfnamefont {B.~P.}\ \bibnamefont
  {Abbott}} \emph {et~al.} (\bibinfo {collaboration} {LIGO Scientific,
  Virgo}),\ }\href {\doibase 10.1088/1361-6382/aaaafa} {\bibfield  {journal}
  {\bibinfo  {journal} {Class. Quant. Grav.}\ }\textbf {\bibinfo {volume}
  {35}},\ \bibinfo {pages} {065010} (\bibinfo {year} {2018}{\natexlab{b}})},\
  \Eprint {http://arxiv.org/abs/1710.02185} {arXiv:1710.02185 [gr-qc]}
  \BibitemShut {NoStop}%
\bibitem [{\citenamefont {Tsukada}\ \emph {et~al.}(2023)\citenamefont {Tsukada}
  \emph {et~al.}}]{Tsukada:2023edh}%
  \BibitemOpen
  \bibfield  {author} {\bibinfo {author} {\bibfnamefont {L.}~\bibnamefont
  {Tsukada}} \emph {et~al.},\ }\href {\doibase 10.1103/PhysRevD.108.043004}
  {\bibfield  {journal} {\bibinfo  {journal} {Phys. Rev. D}\ }\textbf {\bibinfo
  {volume} {108}},\ \bibinfo {pages} {043004} (\bibinfo {year} {2023})},\
  \Eprint {http://arxiv.org/abs/2305.06286} {arXiv:2305.06286 [astro-ph.IM]}
  \BibitemShut {NoStop}%
\bibitem [{\citenamefont {Aubin}\ \emph {et~al.}(2021)\citenamefont {Aubin}
  \emph {et~al.}}]{Aubin:2020goo}%
  \BibitemOpen
  \bibfield  {author} {\bibinfo {author} {\bibfnamefont {F.}~\bibnamefont
  {Aubin}} \emph {et~al.},\ }\href {\doibase 10.1088/1361-6382/abe913}
  {\bibfield  {journal} {\bibinfo  {journal} {Class. Quant. Grav.}\ }\textbf
  {\bibinfo {volume} {38}},\ \bibinfo {pages} {095004} (\bibinfo {year}
  {2021})},\ \Eprint {http://arxiv.org/abs/2012.11512} {arXiv:2012.11512
  [gr-qc]} \BibitemShut {NoStop}%
\bibitem [{\citenamefont {Nitz}\ \emph {et~al.}(2018)\citenamefont {Nitz},
  \citenamefont {Dal~Canton}, \citenamefont {Davis},\ and\ \citenamefont
  {Reyes}}]{Nitz:2018rgo}%
  \BibitemOpen
  \bibfield  {author} {\bibinfo {author} {\bibfnamefont {A.~H.}\ \bibnamefont
  {Nitz}}, \bibinfo {author} {\bibfnamefont {T.}~\bibnamefont {Dal~Canton}},
  \bibinfo {author} {\bibfnamefont {D.}~\bibnamefont {Davis}}, \ and\ \bibinfo
  {author} {\bibfnamefont {S.}~\bibnamefont {Reyes}},\ }\href {\doibase
  10.1103/PhysRevD.98.024050} {\bibfield  {journal} {\bibinfo  {journal} {Phys.
  Rev. D}\ }\textbf {\bibinfo {volume} {98}},\ \bibinfo {pages} {024050}
  (\bibinfo {year} {2018})},\ \Eprint {http://arxiv.org/abs/1805.11174}
  {arXiv:1805.11174 [gr-qc]} \BibitemShut {NoStop}%
\bibitem [{\citenamefont {Cuoco}\ \emph {et~al.}(2021)\citenamefont {Cuoco}
  \emph {et~al.}}]{Cuoco:2020ogp}%
  \BibitemOpen
  \bibfield  {author} {\bibinfo {author} {\bibfnamefont {E.}~\bibnamefont
  {Cuoco}} \emph {et~al.},\ }\href {\doibase 10.1088/2632-2153/abb93a}
  {\bibfield  {journal} {\bibinfo  {journal} {Mach. Learn. Sci. Tech.}\
  }\textbf {\bibinfo {volume} {2}},\ \bibinfo {pages} {011002} (\bibinfo {year}
  {2021})},\ \Eprint {http://arxiv.org/abs/2005.03745} {arXiv:2005.03745
  [astro-ph.HE]} \BibitemShut {NoStop}%
\bibitem [{\citenamefont {Stergioulas}(2024)}]{Stergioulas:2024jgk}%
  \BibitemOpen
  \bibfield  {author} {\bibinfo {author} {\bibfnamefont {N.}~\bibnamefont
  {Stergioulas}}\ }(\bibinfo {year} {2024})\ \Eprint
  {http://arxiv.org/abs/2401.07406} {arXiv:2401.07406 [gr-qc]} \BibitemShut
  {NoStop}%
\bibitem [{\citenamefont {George}\ and\ \citenamefont
  {Huerta}(2018)}]{George:2017pmj}%
  \BibitemOpen
  \bibfield  {author} {\bibinfo {author} {\bibfnamefont {D.}~\bibnamefont
  {George}}\ and\ \bibinfo {author} {\bibfnamefont {E.~A.}\ \bibnamefont
  {Huerta}},\ }\href {\doibase 10.1016/j.physletb.2017.12.053} {\bibfield
  {journal} {\bibinfo  {journal} {Phys. Lett. B}\ }\textbf {\bibinfo {volume}
  {778}},\ \bibinfo {pages} {64} (\bibinfo {year} {2018})},\ \Eprint
  {http://arxiv.org/abs/1711.03121} {arXiv:1711.03121 [gr-qc]} \BibitemShut
  {NoStop}%
\bibitem [{\citenamefont {Gabbard}\ \emph {et~al.}(2018)\citenamefont
  {Gabbard}, \citenamefont {Williams}, \citenamefont {Hayes},\ and\
  \citenamefont {Messenger}}]{Gabbard:2017lja}%
  \BibitemOpen
  \bibfield  {author} {\bibinfo {author} {\bibfnamefont {H.}~\bibnamefont
  {Gabbard}}, \bibinfo {author} {\bibfnamefont {M.}~\bibnamefont {Williams}},
  \bibinfo {author} {\bibfnamefont {F.}~\bibnamefont {Hayes}}, \ and\ \bibinfo
  {author} {\bibfnamefont {C.}~\bibnamefont {Messenger}},\ }\href {\doibase
  10.1103/PhysRevLett.120.141103} {\bibfield  {journal} {\bibinfo  {journal}
  {Phys. Rev. Lett.}\ }\textbf {\bibinfo {volume} {120}},\ \bibinfo {pages}
  {141103} (\bibinfo {year} {2018})},\ \Eprint
  {http://arxiv.org/abs/1712.06041} {arXiv:1712.06041 [astro-ph.IM]}
  \BibitemShut {NoStop}%
\bibitem [{\citenamefont {Nousi}\ \emph {et~al.}(2023)\citenamefont {Nousi},
  \citenamefont {Koloniari}, \citenamefont {Passalis}, \citenamefont {Iosif},
  \citenamefont {Stergioulas},\ and\ \citenamefont {Tefas}}]{Nousi:2022dwh}%
  \BibitemOpen
  \bibfield  {author} {\bibinfo {author} {\bibfnamefont {P.}~\bibnamefont
  {Nousi}}, \bibinfo {author} {\bibfnamefont {A.~E.}\ \bibnamefont
  {Koloniari}}, \bibinfo {author} {\bibfnamefont {N.}~\bibnamefont {Passalis}},
  \bibinfo {author} {\bibfnamefont {P.}~\bibnamefont {Iosif}}, \bibinfo
  {author} {\bibfnamefont {N.}~\bibnamefont {Stergioulas}}, \ and\ \bibinfo
  {author} {\bibfnamefont {A.}~\bibnamefont {Tefas}},\ }\href {\doibase
  10.1103/PhysRevD.108.024022} {\bibfield  {journal} {\bibinfo  {journal}
  {Phys. Rev. D}\ }\textbf {\bibinfo {volume} {108}},\ \bibinfo {pages}
  {024022} (\bibinfo {year} {2023})},\ \Eprint
  {http://arxiv.org/abs/2211.01520} {arXiv:2211.01520 [gr-qc]} \BibitemShut
  {NoStop}%
\bibitem [{\citenamefont {Marx}\ \emph {et~al.}(2024)\citenamefont {Marx} \emph
  {et~al.}}]{Marx:2024wjt}%
  \BibitemOpen
  \bibfield  {author} {\bibinfo {author} {\bibfnamefont {E.}~\bibnamefont
  {Marx}} \emph {et~al.},\ }\href@noop {} {\  (\bibinfo {year} {2024})},\
  \Eprint {http://arxiv.org/abs/2403.18661} {arXiv:2403.18661 [gr-qc]}
  \BibitemShut {NoStop}%
\bibitem [{\citenamefont {Alfaidi}\ and\ \citenamefont
  {Messenger}(2024)}]{Alfaidi:2024ioo}%
  \BibitemOpen
  \bibfield  {author} {\bibinfo {author} {\bibfnamefont {R.}~\bibnamefont
  {Alfaidi}}\ and\ \bibinfo {author} {\bibfnamefont {C.}~\bibnamefont
  {Messenger}},\ }\href@noop {} {\  (\bibinfo {year} {2024})},\ \Eprint
  {http://arxiv.org/abs/2402.04589} {arXiv:2402.04589 [gr-qc]} \BibitemShut
  {NoStop}%
\bibitem [{\citenamefont {Sch\"afer}\ \emph {et~al.}(2023)\citenamefont
  {Sch\"afer} \emph {et~al.}}]{Schafer:2022dxv}%
  \BibitemOpen
  \bibfield  {author} {\bibinfo {author} {\bibfnamefont {M.~B.}\ \bibnamefont
  {Sch\"afer}} \emph {et~al.},\ }\href {\doibase 10.1103/PhysRevD.107.023021}
  {\bibfield  {journal} {\bibinfo  {journal} {Phys. Rev. D}\ }\textbf {\bibinfo
  {volume} {107}},\ \bibinfo {pages} {023021} (\bibinfo {year} {2023})},\
  \Eprint {http://arxiv.org/abs/2209.11146} {arXiv:2209.11146 [astro-ph.IM]}
  \BibitemShut {NoStop}%
\bibitem [{\citenamefont {Mishra}\ \emph {et~al.}(2022)\citenamefont {Mishra}
  \emph {et~al.}}]{Mishra:2022ott}%
  \BibitemOpen
  \bibfield  {author} {\bibinfo {author} {\bibfnamefont {T.}~\bibnamefont
  {Mishra}} \emph {et~al.},\ }\href {\doibase 10.1103/PhysRevD.105.083018}
  {\bibfield  {journal} {\bibinfo  {journal} {Phys. Rev. D}\ }\textbf {\bibinfo
  {volume} {105}},\ \bibinfo {pages} {083018} (\bibinfo {year} {2022})},\
  \Eprint {http://arxiv.org/abs/2201.01495} {arXiv:2201.01495 [gr-qc]}
  \BibitemShut {NoStop}%
\bibitem [{\citenamefont {Krastev}(2020)}]{Krastev:2019koe}%
  \BibitemOpen
  \bibfield  {author} {\bibinfo {author} {\bibfnamefont {P.~G.}\ \bibnamefont
  {Krastev}},\ }\href {\doibase 10.1016/j.physletb.2020.135330} {\bibfield
  {journal} {\bibinfo  {journal} {Phys. Lett. B}\ }\textbf {\bibinfo {volume}
  {803}},\ \bibinfo {pages} {135330} (\bibinfo {year} {2020})},\ \Eprint
  {http://arxiv.org/abs/1908.03151} {arXiv:1908.03151 [astro-ph.IM]}
  \BibitemShut {NoStop}%
\bibitem [{\citenamefont {Sch\"afer}\ \emph {et~al.}(2020)\citenamefont
  {Sch\"afer}, \citenamefont {Ohme},\ and\ \citenamefont
  {Nitz}}]{Schafer:2020kor}%
  \BibitemOpen
  \bibfield  {author} {\bibinfo {author} {\bibfnamefont {M.~B.}\ \bibnamefont
  {Sch\"afer}}, \bibinfo {author} {\bibfnamefont {F.}~\bibnamefont {Ohme}}, \
  and\ \bibinfo {author} {\bibfnamefont {A.~H.}\ \bibnamefont {Nitz}},\ }\href
  {\doibase 10.1103/PhysRevD.102.063015} {\bibfield  {journal} {\bibinfo
  {journal} {Phys. Rev. D}\ }\textbf {\bibinfo {volume} {102}},\ \bibinfo
  {pages} {063015} (\bibinfo {year} {2020})},\ \Eprint
  {http://arxiv.org/abs/2006.01509} {arXiv:2006.01509 [astro-ph.HE]}
  \BibitemShut {NoStop}%
\bibitem [{\citenamefont {Wei}\ and\ \citenamefont
  {Huerta}(2021)}]{Wei:2020sfz}%
  \BibitemOpen
  \bibfield  {author} {\bibinfo {author} {\bibfnamefont {W.}~\bibnamefont
  {Wei}}\ and\ \bibinfo {author} {\bibfnamefont {E.~A.}\ \bibnamefont
  {Huerta}},\ }\href {\doibase 10.1016/j.physletb.2021.136185} {\bibfield
  {journal} {\bibinfo  {journal} {Phys. Lett. B}\ }\textbf {\bibinfo {volume}
  {816}},\ \bibinfo {pages} {136185} (\bibinfo {year} {2021})},\ \Eprint
  {http://arxiv.org/abs/2010.09751} {arXiv:2010.09751 [gr-qc]} \BibitemShut
  {NoStop}%
\bibitem [{\citenamefont {Baltus}\ \emph {et~al.}(2021)\citenamefont {Baltus},
  \citenamefont {Janquart}, \citenamefont {Lopez}, \citenamefont {Reza},
  \citenamefont {Caudill},\ and\ \citenamefont {Cudell}}]{Baltus:2021nme}%
  \BibitemOpen
  \bibfield  {author} {\bibinfo {author} {\bibfnamefont {G.}~\bibnamefont
  {Baltus}}, \bibinfo {author} {\bibfnamefont {J.}~\bibnamefont {Janquart}},
  \bibinfo {author} {\bibfnamefont {M.}~\bibnamefont {Lopez}}, \bibinfo
  {author} {\bibfnamefont {A.}~\bibnamefont {Reza}}, \bibinfo {author}
  {\bibfnamefont {S.}~\bibnamefont {Caudill}}, \ and\ \bibinfo {author}
  {\bibfnamefont {J.-R.}\ \bibnamefont {Cudell}},\ }\href {\doibase
  10.1103/PhysRevD.103.102003} {\bibfield  {journal} {\bibinfo  {journal}
  {Phys. Rev. D}\ }\textbf {\bibinfo {volume} {103}},\ \bibinfo {pages}
  {102003} (\bibinfo {year} {2021})},\ \Eprint
  {http://arxiv.org/abs/2104.00594} {arXiv:2104.00594 [gr-qc]} \BibitemShut
  {NoStop}%
\bibitem [{\citenamefont {Yu}\ \emph {et~al.}(2021)\citenamefont {Yu},
  \citenamefont {Adhikari}, \citenamefont {Magee}, \citenamefont {Sachdev},\
  and\ \citenamefont {Chen}}]{Yu:2021vvm}%
  \BibitemOpen
  \bibfield  {author} {\bibinfo {author} {\bibfnamefont {H.}~\bibnamefont
  {Yu}}, \bibinfo {author} {\bibfnamefont {R.~X.}\ \bibnamefont {Adhikari}},
  \bibinfo {author} {\bibfnamefont {R.}~\bibnamefont {Magee}}, \bibinfo
  {author} {\bibfnamefont {S.}~\bibnamefont {Sachdev}}, \ and\ \bibinfo
  {author} {\bibfnamefont {Y.}~\bibnamefont {Chen}},\ }\href {\doibase
  10.1103/PhysRevD.104.062004} {\bibfield  {journal} {\bibinfo  {journal}
  {Phys. Rev. D}\ }\textbf {\bibinfo {volume} {104}},\ \bibinfo {pages}
  {062004} (\bibinfo {year} {2021})},\ \Eprint
  {http://arxiv.org/abs/2104.09438} {arXiv:2104.09438 [gr-qc]} \BibitemShut
  {NoStop}%
\bibitem [{\citenamefont {Baltus}\ \emph {et~al.}(2022)\citenamefont {Baltus},
  \citenamefont {Janquart}, \citenamefont {Lopez}, \citenamefont {Narola},\
  and\ \citenamefont {Cudell}}]{Baltus:2022pep}%
  \BibitemOpen
  \bibfield  {author} {\bibinfo {author} {\bibfnamefont {G.}~\bibnamefont
  {Baltus}}, \bibinfo {author} {\bibfnamefont {J.}~\bibnamefont {Janquart}},
  \bibinfo {author} {\bibfnamefont {M.}~\bibnamefont {Lopez}}, \bibinfo
  {author} {\bibfnamefont {H.}~\bibnamefont {Narola}}, \ and\ \bibinfo {author}
  {\bibfnamefont {J.-R.}\ \bibnamefont {Cudell}},\ }\href {\doibase
  10.1103/PhysRevD.106.042002} {\bibfield  {journal} {\bibinfo  {journal}
  {Phys. Rev. D}\ }\textbf {\bibinfo {volume} {106}},\ \bibinfo {pages}
  {042002} (\bibinfo {year} {2022})},\ \Eprint
  {http://arxiv.org/abs/2205.04750} {arXiv:2205.04750 [gr-qc]} \BibitemShut
  {NoStop}%
\bibitem [{\citenamefont {Astone}\ \emph {et~al.}(2018)\citenamefont {Astone},
  \citenamefont {Cerd\'a-Dur\'an}, \citenamefont {Di~Palma}, \citenamefont
  {Drago}, \citenamefont {Muciaccia}, \citenamefont {Palomba},\ and\
  \citenamefont {Ricci}}]{Astone:2018uge}%
  \BibitemOpen
  \bibfield  {author} {\bibinfo {author} {\bibfnamefont {P.}~\bibnamefont
  {Astone}}, \bibinfo {author} {\bibfnamefont {P.}~\bibnamefont
  {Cerd\'a-Dur\'an}}, \bibinfo {author} {\bibfnamefont {I.}~\bibnamefont
  {Di~Palma}}, \bibinfo {author} {\bibfnamefont {M.}~\bibnamefont {Drago}},
  \bibinfo {author} {\bibfnamefont {F.}~\bibnamefont {Muciaccia}}, \bibinfo
  {author} {\bibfnamefont {C.}~\bibnamefont {Palomba}}, \ and\ \bibinfo
  {author} {\bibfnamefont {F.}~\bibnamefont {Ricci}},\ }\href {\doibase
  10.1103/PhysRevD.98.122002} {\bibfield  {journal} {\bibinfo  {journal} {Phys.
  Rev. D}\ }\textbf {\bibinfo {volume} {98}},\ \bibinfo {pages} {122002}
  (\bibinfo {year} {2018})},\ \Eprint {http://arxiv.org/abs/1812.05363}
  {arXiv:1812.05363 [astro-ph.IM]} \BibitemShut {NoStop}%
\bibitem [{\citenamefont {Antelis}\ \emph {et~al.}(2022)\citenamefont
  {Antelis}, \citenamefont {Cavaglia}, \citenamefont {Hansen}, \citenamefont
  {Morales}, \citenamefont {Moreno}, \citenamefont {Mukherjee}, \citenamefont
  {Szczepa\'nczyk},\ and\ \citenamefont {Zanolin}}]{Antelis:2021qak}%
  \BibitemOpen
  \bibfield  {author} {\bibinfo {author} {\bibfnamefont {J.~M.}\ \bibnamefont
  {Antelis}}, \bibinfo {author} {\bibfnamefont {M.}~\bibnamefont {Cavaglia}},
  \bibinfo {author} {\bibfnamefont {T.}~\bibnamefont {Hansen}}, \bibinfo
  {author} {\bibfnamefont {M.~D.}\ \bibnamefont {Morales}}, \bibinfo {author}
  {\bibfnamefont {C.}~\bibnamefont {Moreno}}, \bibinfo {author} {\bibfnamefont
  {S.}~\bibnamefont {Mukherjee}}, \bibinfo {author} {\bibfnamefont {M.~J.}\
  \bibnamefont {Szczepa\'nczyk}}, \ and\ \bibinfo {author} {\bibfnamefont
  {M.}~\bibnamefont {Zanolin}},\ }\href {\doibase 10.1103/PhysRevD.105.084054}
  {\bibfield  {journal} {\bibinfo  {journal} {Phys. Rev. D}\ }\textbf {\bibinfo
  {volume} {105}},\ \bibinfo {pages} {084054} (\bibinfo {year} {2022})},\
  \Eprint {http://arxiv.org/abs/2111.07219} {arXiv:2111.07219 [gr-qc]}
  \BibitemShut {NoStop}%
\bibitem [{\citenamefont {L\'opez~Portilla}\ \emph {et~al.}(2021)\citenamefont
  {L\'opez~Portilla}, \citenamefont {Palma}, \citenamefont {Drago},
  \citenamefont {Cerd\'a-Dur\'an},\ and\ \citenamefont
  {Ricci}}]{LopezPortilla:2020odz}%
  \BibitemOpen
  \bibfield  {author} {\bibinfo {author} {\bibfnamefont {M.}~\bibnamefont
  {L\'opez~Portilla}}, \bibinfo {author} {\bibfnamefont {I.~D.}\ \bibnamefont
  {Palma}}, \bibinfo {author} {\bibfnamefont {M.}~\bibnamefont {Drago}},
  \bibinfo {author} {\bibfnamefont {P.}~\bibnamefont {Cerd\'a-Dur\'an}}, \ and\
  \bibinfo {author} {\bibfnamefont {F.}~\bibnamefont {Ricci}},\ }\href
  {\doibase 10.1103/PhysRevD.103.063011} {\bibfield  {journal} {\bibinfo
  {journal} {Phys. Rev. D}\ }\textbf {\bibinfo {volume} {103}},\ \bibinfo
  {pages} {063011} (\bibinfo {year} {2021})},\ \Eprint
  {http://arxiv.org/abs/2011.13733} {arXiv:2011.13733 [astro-ph.IM]}
  \BibitemShut {NoStop}%
\bibitem [{\citenamefont {Gabbard}\ \emph {et~al.}(2022)\citenamefont
  {Gabbard}, \citenamefont {Messenger}, \citenamefont {Heng}, \citenamefont
  {Tonolini},\ and\ \citenamefont {Murray-Smith}}]{Gabbard:2019rde}%
  \BibitemOpen
  \bibfield  {author} {\bibinfo {author} {\bibfnamefont {H.}~\bibnamefont
  {Gabbard}}, \bibinfo {author} {\bibfnamefont {C.}~\bibnamefont {Messenger}},
  \bibinfo {author} {\bibfnamefont {I.~S.}\ \bibnamefont {Heng}}, \bibinfo
  {author} {\bibfnamefont {F.}~\bibnamefont {Tonolini}}, \ and\ \bibinfo
  {author} {\bibfnamefont {R.}~\bibnamefont {Murray-Smith}},\ }\href {\doibase
  10.1038/s41567-021-01425-7} {\bibfield  {journal} {\bibinfo  {journal}
  {Nature Phys.}\ }\textbf {\bibinfo {volume} {18}},\ \bibinfo {pages} {112}
  (\bibinfo {year} {2022})},\ \Eprint {http://arxiv.org/abs/1909.06296}
  {arXiv:1909.06296 [astro-ph.IM]} \BibitemShut {NoStop}%
\bibitem [{\citenamefont {Dax}\ \emph {et~al.}(2021)\citenamefont {Dax},
  \citenamefont {Green}, \citenamefont {Gair}, \citenamefont {Macke},
  \citenamefont {Buonanno},\ and\ \citenamefont {Sch\"olkopf}}]{Dax:2021tsq}%
  \BibitemOpen
  \bibfield  {author} {\bibinfo {author} {\bibfnamefont {M.}~\bibnamefont
  {Dax}}, \bibinfo {author} {\bibfnamefont {S.~R.}\ \bibnamefont {Green}},
  \bibinfo {author} {\bibfnamefont {J.}~\bibnamefont {Gair}}, \bibinfo {author}
  {\bibfnamefont {J.~H.}\ \bibnamefont {Macke}}, \bibinfo {author}
  {\bibfnamefont {A.}~\bibnamefont {Buonanno}}, \ and\ \bibinfo {author}
  {\bibfnamefont {B.}~\bibnamefont {Sch\"olkopf}},\ }\href {\doibase
  10.1103/PhysRevLett.127.241103} {\bibfield  {journal} {\bibinfo  {journal}
  {Phys. Rev. Lett.}\ }\textbf {\bibinfo {volume} {127}},\ \bibinfo {pages}
  {241103} (\bibinfo {year} {2021})},\ \Eprint
  {http://arxiv.org/abs/2106.12594} {arXiv:2106.12594 [gr-qc]} \BibitemShut
  {NoStop}%
\bibitem [{\citenamefont {Chatterjee}\ \emph {et~al.}(2022)\citenamefont
  {Chatterjee}, \citenamefont {Wen}, \citenamefont {Beveridge}, \citenamefont
  {Diakogiannis},\ and\ \citenamefont {Vinsen}}]{Chatterjee:2022ggk}%
  \BibitemOpen
  \bibfield  {author} {\bibinfo {author} {\bibfnamefont {C.}~\bibnamefont
  {Chatterjee}}, \bibinfo {author} {\bibfnamefont {L.}~\bibnamefont {Wen}},
  \bibinfo {author} {\bibfnamefont {D.}~\bibnamefont {Beveridge}}, \bibinfo
  {author} {\bibfnamefont {F.}~\bibnamefont {Diakogiannis}}, \ and\ \bibinfo
  {author} {\bibfnamefont {K.}~\bibnamefont {Vinsen}},\ }\href@noop {} {\
  (\bibinfo {year} {2022})},\ \Eprint {http://arxiv.org/abs/2207.14522}
  {arXiv:2207.14522 [gr-qc]} \BibitemShut {NoStop}%
\bibitem [{\citenamefont {Chatterjee}\ and\ \citenamefont
  {Wen}(2022)}]{Chatterjee:2022dik}%
  \BibitemOpen
  \bibfield  {author} {\bibinfo {author} {\bibfnamefont {C.}~\bibnamefont
  {Chatterjee}}\ and\ \bibinfo {author} {\bibfnamefont {L.}~\bibnamefont
  {Wen}},\ }\href@noop {} {\  (\bibinfo {year} {2022})},\ \Eprint
  {http://arxiv.org/abs/2301.03558} {arXiv:2301.03558 [astro-ph.HE]}
  \BibitemShut {NoStop}%
\bibitem [{\citenamefont {Dax}\ \emph {et~al.}(2023)\citenamefont {Dax},
  \citenamefont {Green}, \citenamefont {Gair}, \citenamefont {P\"urrer},
  \citenamefont {Wildberger}, \citenamefont {Macke}, \citenamefont {Buonanno},\
  and\ \citenamefont {Sch\"olkopf}}]{Dax:2022pxd}%
  \BibitemOpen
  \bibfield  {author} {\bibinfo {author} {\bibfnamefont {M.}~\bibnamefont
  {Dax}}, \bibinfo {author} {\bibfnamefont {S.~R.}\ \bibnamefont {Green}},
  \bibinfo {author} {\bibfnamefont {J.}~\bibnamefont {Gair}}, \bibinfo {author}
  {\bibfnamefont {M.}~\bibnamefont {P\"urrer}}, \bibinfo {author}
  {\bibfnamefont {J.}~\bibnamefont {Wildberger}}, \bibinfo {author}
  {\bibfnamefont {J.~H.}\ \bibnamefont {Macke}}, \bibinfo {author}
  {\bibfnamefont {A.}~\bibnamefont {Buonanno}}, \ and\ \bibinfo {author}
  {\bibfnamefont {B.}~\bibnamefont {Sch\"olkopf}},\ }\href {\doibase
  10.1103/PhysRevLett.130.171403} {\bibfield  {journal} {\bibinfo  {journal}
  {Phys. Rev. Lett.}\ }\textbf {\bibinfo {volume} {130}},\ \bibinfo {pages}
  {171403} (\bibinfo {year} {2023})},\ \Eprint
  {http://arxiv.org/abs/2210.05686} {arXiv:2210.05686 [gr-qc]} \BibitemShut
  {NoStop}%
\bibitem [{\citenamefont {Wong}\ \emph {et~al.}(2023)\citenamefont {Wong},
  \citenamefont {Isi},\ and\ \citenamefont {Edwards}}]{Wong:2023lgb}%
  \BibitemOpen
  \bibfield  {author} {\bibinfo {author} {\bibfnamefont {K.~W.~K.}\
  \bibnamefont {Wong}}, \bibinfo {author} {\bibfnamefont {M.}~\bibnamefont
  {Isi}}, \ and\ \bibinfo {author} {\bibfnamefont {T.~D.~P.}\ \bibnamefont
  {Edwards}},\ }\href@noop {} {\  (\bibinfo {year} {2023})},\ \Eprint
  {http://arxiv.org/abs/2302.05333} {arXiv:2302.05333 [astro-ph.IM]}
  \BibitemShut {NoStop}%
\bibitem [{\citenamefont {Schmidt}\ \emph {et~al.}(2021)\citenamefont
  {Schmidt}, \citenamefont {Breschi}, \citenamefont {Gamba}, \citenamefont
  {Pagano}, \citenamefont {Rettegno}, \citenamefont {Riemenschneider},
  \citenamefont {Bernuzzi}, \citenamefont {Nagar},\ and\ \citenamefont
  {Del~Pozzo}}]{Schmidt:2020yuu}%
  \BibitemOpen
  \bibfield  {author} {\bibinfo {author} {\bibfnamefont {S.}~\bibnamefont
  {Schmidt}}, \bibinfo {author} {\bibfnamefont {M.}~\bibnamefont {Breschi}},
  \bibinfo {author} {\bibfnamefont {R.}~\bibnamefont {Gamba}}, \bibinfo
  {author} {\bibfnamefont {G.}~\bibnamefont {Pagano}}, \bibinfo {author}
  {\bibfnamefont {P.}~\bibnamefont {Rettegno}}, \bibinfo {author}
  {\bibfnamefont {G.}~\bibnamefont {Riemenschneider}}, \bibinfo {author}
  {\bibfnamefont {S.}~\bibnamefont {Bernuzzi}}, \bibinfo {author}
  {\bibfnamefont {A.}~\bibnamefont {Nagar}}, \ and\ \bibinfo {author}
  {\bibfnamefont {W.}~\bibnamefont {Del~Pozzo}},\ }\href {\doibase
  10.1103/PhysRevD.103.043020} {\bibfield  {journal} {\bibinfo  {journal}
  {Phys. Rev. D}\ }\textbf {\bibinfo {volume} {103}},\ \bibinfo {pages}
  {043020} (\bibinfo {year} {2021})},\ \Eprint
  {http://arxiv.org/abs/2011.01958} {arXiv:2011.01958 [gr-qc]} \BibitemShut
  {NoStop}%
\bibitem [{\citenamefont {Chua}\ \emph {et~al.}(2019)\citenamefont {Chua},
  \citenamefont {Galley},\ and\ \citenamefont {Vallisneri}}]{Chua:2018woh}%
  \BibitemOpen
  \bibfield  {author} {\bibinfo {author} {\bibfnamefont {A.~J.~K.}\
  \bibnamefont {Chua}}, \bibinfo {author} {\bibfnamefont {C.~R.}\ \bibnamefont
  {Galley}}, \ and\ \bibinfo {author} {\bibfnamefont {M.}~\bibnamefont
  {Vallisneri}},\ }\href {\doibase 10.1103/PhysRevLett.122.211101} {\bibfield
  {journal} {\bibinfo  {journal} {Phys. Rev. Lett.}\ }\textbf {\bibinfo
  {volume} {122}},\ \bibinfo {pages} {211101} (\bibinfo {year} {2019})},\
  \Eprint {http://arxiv.org/abs/1811.05491} {arXiv:1811.05491 [astro-ph.IM]}
  \BibitemShut {NoStop}%
\bibitem [{\citenamefont {Khan}\ and\ \citenamefont
  {Green}(2021)}]{Khan:2020fso}%
  \BibitemOpen
  \bibfield  {author} {\bibinfo {author} {\bibfnamefont {S.}~\bibnamefont
  {Khan}}\ and\ \bibinfo {author} {\bibfnamefont {R.}~\bibnamefont {Green}},\
  }\href {\doibase 10.1103/PhysRevD.103.064015} {\bibfield  {journal} {\bibinfo
   {journal} {Phys. Rev. D}\ }\textbf {\bibinfo {volume} {103}},\ \bibinfo
  {pages} {064015} (\bibinfo {year} {2021})},\ \Eprint
  {http://arxiv.org/abs/2008.12932} {arXiv:2008.12932 [gr-qc]} \BibitemShut
  {NoStop}%
\bibitem [{\citenamefont {Tissino}\ \emph {et~al.}(2023)\citenamefont
  {Tissino}, \citenamefont {Carullo}, \citenamefont {Breschi}, \citenamefont
  {Gamba}, \citenamefont {Schmidt},\ and\ \citenamefont
  {Bernuzzi}}]{Tissino:2022thn}%
  \BibitemOpen
  \bibfield  {author} {\bibinfo {author} {\bibfnamefont {J.}~\bibnamefont
  {Tissino}}, \bibinfo {author} {\bibfnamefont {G.}~\bibnamefont {Carullo}},
  \bibinfo {author} {\bibfnamefont {M.}~\bibnamefont {Breschi}}, \bibinfo
  {author} {\bibfnamefont {R.}~\bibnamefont {Gamba}}, \bibinfo {author}
  {\bibfnamefont {S.}~\bibnamefont {Schmidt}}, \ and\ \bibinfo {author}
  {\bibfnamefont {S.}~\bibnamefont {Bernuzzi}},\ }\href {\doibase
  10.1103/PhysRevD.107.084037} {\bibfield  {journal} {\bibinfo  {journal}
  {Phys. Rev. D}\ }\textbf {\bibinfo {volume} {107}},\ \bibinfo {pages}
  {084037} (\bibinfo {year} {2023})},\ \Eprint
  {http://arxiv.org/abs/2210.15684} {arXiv:2210.15684 [gr-qc]} \BibitemShut
  {NoStop}%
\bibitem [{\citenamefont {Thomas}\ \emph {et~al.}(2022)\citenamefont {Thomas},
  \citenamefont {Pratten},\ and\ \citenamefont {Schmidt}}]{Thomas:2022rmc}%
  \BibitemOpen
  \bibfield  {author} {\bibinfo {author} {\bibfnamefont {L.~M.}\ \bibnamefont
  {Thomas}}, \bibinfo {author} {\bibfnamefont {G.}~\bibnamefont {Pratten}}, \
  and\ \bibinfo {author} {\bibfnamefont {P.}~\bibnamefont {Schmidt}},\ }\href
  {\doibase 10.1103/PhysRevD.106.104029} {\bibfield  {journal} {\bibinfo
  {journal} {Phys. Rev. D}\ }\textbf {\bibinfo {volume} {106}},\ \bibinfo
  {pages} {104029} (\bibinfo {year} {2022})},\ \Eprint
  {http://arxiv.org/abs/2205.14066} {arXiv:2205.14066 [gr-qc]} \BibitemShut
  {NoStop}%
\bibitem [{\citenamefont {Magee}\ \emph
  {et~al.}(2024{\natexlab{b}})\citenamefont {Magee}, \citenamefont {George},
  \citenamefont {Li},\ and\ \citenamefont {Sharma}}]{Magee:2024yal}%
  \BibitemOpen
  \bibfield  {author} {\bibinfo {author} {\bibfnamefont {R.}~\bibnamefont
  {Magee}}, \bibinfo {author} {\bibfnamefont {R.}~\bibnamefont {George}},
  \bibinfo {author} {\bibfnamefont {A.}~\bibnamefont {Li}}, \ and\ \bibinfo
  {author} {\bibfnamefont {R.}~\bibnamefont {Sharma}},\ }\href@noop {} {\
  (\bibinfo {year} {2024}{\natexlab{b}})},\ \Eprint
  {http://arxiv.org/abs/2408.02470} {arXiv:2408.02470 [astro-ph.IM]}
  \BibitemShut {NoStop}%
\bibitem [{\citenamefont {Vajente}\ \emph {et~al.}(2020)\citenamefont
  {Vajente}, \citenamefont {Huang}, \citenamefont {Isi}, \citenamefont
  {Driggers}, \citenamefont {Kissel}, \citenamefont {Szczepanczyk},\ and\
  \citenamefont {Vitale}}]{Vajente:2019ycy}%
  \BibitemOpen
  \bibfield  {author} {\bibinfo {author} {\bibfnamefont {G.}~\bibnamefont
  {Vajente}}, \bibinfo {author} {\bibfnamefont {Y.}~\bibnamefont {Huang}},
  \bibinfo {author} {\bibfnamefont {M.}~\bibnamefont {Isi}}, \bibinfo {author}
  {\bibfnamefont {J.~C.}\ \bibnamefont {Driggers}}, \bibinfo {author}
  {\bibfnamefont {J.~S.}\ \bibnamefont {Kissel}}, \bibinfo {author}
  {\bibfnamefont {M.~J.}\ \bibnamefont {Szczepanczyk}}, \ and\ \bibinfo
  {author} {\bibfnamefont {S.}~\bibnamefont {Vitale}},\ }\href {\doibase
  10.1103/PhysRevD.101.042003} {\bibfield  {journal} {\bibinfo  {journal}
  {Phys. Rev. D}\ }\textbf {\bibinfo {volume} {101}},\ \bibinfo {pages}
  {042003} (\bibinfo {year} {2020})},\ \Eprint
  {http://arxiv.org/abs/1911.09083} {arXiv:1911.09083 [gr-qc]} \BibitemShut
  {NoStop}%
\bibitem [{\citenamefont {Essick}\ \emph {et~al.}(2020)\citenamefont {Essick},
  \citenamefont {Godwin}, \citenamefont {Hanna}, \citenamefont {Blackburn},\
  and\ \citenamefont {Katsavounidis}}]{Essick:2020qpo}%
  \BibitemOpen
  \bibfield  {author} {\bibinfo {author} {\bibfnamefont {R.}~\bibnamefont
  {Essick}}, \bibinfo {author} {\bibfnamefont {P.}~\bibnamefont {Godwin}},
  \bibinfo {author} {\bibfnamefont {C.}~\bibnamefont {Hanna}}, \bibinfo
  {author} {\bibfnamefont {L.}~\bibnamefont {Blackburn}}, \ and\ \bibinfo
  {author} {\bibfnamefont {E.}~\bibnamefont {Katsavounidis}},\ }\href@noop {}
  {\  (\bibinfo {year} {2020})},\ \Eprint {http://arxiv.org/abs/2005.12761}
  {arXiv:2005.12761 [astro-ph.IM]} \BibitemShut {NoStop}%
\bibitem [{\citenamefont {Saleem}\ \emph {et~al.}(2023)\citenamefont {Saleem}
  \emph {et~al.}}]{Saleem:2023hcm}%
  \BibitemOpen
  \bibfield  {author} {\bibinfo {author} {\bibfnamefont {M.}~\bibnamefont
  {Saleem}} \emph {et~al.},\ }\href@noop {} {\  (\bibinfo {year} {2023})},\
  \Eprint {http://arxiv.org/abs/2306.11366} {arXiv:2306.11366 [gr-qc]}
  \BibitemShut {NoStop}%
\bibitem [{\citenamefont {Zevin}\ \emph {et~al.}(2017)\citenamefont {Zevin}
  \emph {et~al.}}]{Zevin:2016qwy}%
  \BibitemOpen
  \bibfield  {author} {\bibinfo {author} {\bibfnamefont {M.}~\bibnamefont
  {Zevin}} \emph {et~al.},\ }\href {\doibase 10.1088/1361-6382/aa5cea}
  {\bibfield  {journal} {\bibinfo  {journal} {Class. Quant. Grav.}\ }\textbf
  {\bibinfo {volume} {34}},\ \bibinfo {pages} {064003} (\bibinfo {year}
  {2017})},\ \Eprint {http://arxiv.org/abs/1611.04596} {arXiv:1611.04596
  [gr-qc]} \BibitemShut {NoStop}%
\bibitem [{\citenamefont {Glanzer}\ \emph {et~al.}(2023)\citenamefont {Glanzer}
  \emph {et~al.}}]{Glanzer:2022avx}%
  \BibitemOpen
  \bibfield  {author} {\bibinfo {author} {\bibfnamefont {J.}~\bibnamefont
  {Glanzer}} \emph {et~al.},\ }\href {\doibase 10.1088/1361-6382/acb633}
  {\bibfield  {journal} {\bibinfo  {journal} {Class. Quant. Grav.}\ }\textbf
  {\bibinfo {volume} {40}},\ \bibinfo {pages} {065004} (\bibinfo {year}
  {2023})},\ \Eprint {http://arxiv.org/abs/2208.12849} {arXiv:2208.12849
  [gr-qc]} \BibitemShut {NoStop}%
\bibitem [{\citenamefont {Merritt}\ \emph {et~al.}(2021)\citenamefont
  {Merritt}, \citenamefont {Farr} \emph {et~al.}}]{Merritt:2021orq}%
  \BibitemOpen
  \bibfield  {author} {\bibinfo {author} {\bibfnamefont {J.~D.}\ \bibnamefont
  {Merritt}}, \bibinfo {author} {\bibfnamefont {B.}~\bibnamefont {Farr}},
  \emph {et~al.},\ }\href {\doibase 0.1103/PhysRevD.104.102004} {\bibfield
  {journal} {\bibinfo  {journal} {Phys. Rev. D}\ }\textbf {\bibinfo {volume}
  {104}},\ \bibinfo {pages} {11} (\bibinfo {year} {2021})},\ \Eprint
  {http://arxiv.org/abs/2108.12044} {arXiv:2108.12044 [gr-qc]} \BibitemShut
  {NoStop}%
\bibitem [{\citenamefont {{Lopez}}\ \emph
  {et~al.}(2022{\natexlab{a}})\citenamefont {{Lopez}}, \citenamefont
  {{Boudart}}, \citenamefont {{Buijsman}}, \citenamefont {{Reza}},\ and\
  \citenamefont {{Caudill}}}]{2022PhRvD.106b3027L}%
  \BibitemOpen
  \bibfield  {author} {\bibinfo {author} {\bibfnamefont {M.}~\bibnamefont
  {{Lopez}}}, \bibinfo {author} {\bibfnamefont {V.}~\bibnamefont {{Boudart}}},
  \bibinfo {author} {\bibfnamefont {K.}~\bibnamefont {{Buijsman}}}, \bibinfo
  {author} {\bibfnamefont {A.}~\bibnamefont {{Reza}}}, \ and\ \bibinfo {author}
  {\bibfnamefont {S.}~\bibnamefont {{Caudill}}},\ }\href {\doibase
  10.1103/PhysRevD.106.023027} {\bibfield  {journal} {\bibinfo  {journal}
  {\prd}\ }\textbf {\bibinfo {volume} {106}},\ \bibinfo {eid} {023027}
  (\bibinfo {year} {2022}{\natexlab{a}})},\ \Eprint
  {http://arxiv.org/abs/2203.06494} {arXiv:2203.06494 [astro-ph.IM]}
  \BibitemShut {NoStop}%
\bibitem [{\citenamefont {{Lopez}}\ \emph
  {et~al.}(2022{\natexlab{b}})\citenamefont {{Lopez}}, \citenamefont
  {{Boudart}}, \citenamefont {{Schmidt}},\ and\ \citenamefont
  {{Caudill}}}]{2022arXiv220509204L}%
  \BibitemOpen
  \bibfield  {author} {\bibinfo {author} {\bibfnamefont {M.}~\bibnamefont
  {{Lopez}}}, \bibinfo {author} {\bibfnamefont {V.}~\bibnamefont {{Boudart}}},
  \bibinfo {author} {\bibfnamefont {S.}~\bibnamefont {{Schmidt}}}, \ and\
  \bibinfo {author} {\bibfnamefont {S.}~\bibnamefont {{Caudill}}},\ }\href
  {\doibase 10.48550/arXiv.2205.09204} {\bibfield  {journal} {\bibinfo
  {journal} {arXiv e-prints}\ ,\ \bibinfo {eid} {arXiv:2205.09204}} (\bibinfo
  {year} {2022}{\natexlab{b}})},\ \Eprint {http://arxiv.org/abs/2205.09204}
  {arXiv:2205.09204 [astro-ph.IM]} \BibitemShut {NoStop}%
\bibitem [{\citenamefont {Cabero}\ \emph {et~al.}(2019)\citenamefont {Cabero}
  \emph {et~al.}}]{Cabero:2019orq}%
  \BibitemOpen
  \bibfield  {author} {\bibinfo {author} {\bibfnamefont {M.}~\bibnamefont
  {Cabero}} \emph {et~al.},\ }\href {\doibase 10.1088/1361-6382/ab2e14}
  {\bibfield  {journal} {\bibinfo  {journal} {Class. Quant. Grav.}\ }\textbf
  {\bibinfo {volume} {36}},\ \bibinfo {pages} {15} (\bibinfo {year} {2019})},\
  \Eprint {http://arxiv.org/abs/1901.05093} {arXiv:1901.05093
  [physics.ins-det]} \BibitemShut {NoStop}%
\bibitem [{\citenamefont {Davis}\ \emph {et~al.}(2020)\citenamefont {Davis},
  \citenamefont {White},\ and\ \citenamefont {Saulson}}]{Davis:2020nyf}%
  \BibitemOpen
  \bibfield  {author} {\bibinfo {author} {\bibfnamefont {D.}~\bibnamefont
  {Davis}}, \bibinfo {author} {\bibfnamefont {L.~V.}\ \bibnamefont {White}}, \
  and\ \bibinfo {author} {\bibfnamefont {P.~R.}\ \bibnamefont {Saulson}},\
  }\href {\doibase 10.1088/1361-6382/ab91e6} {\bibfield  {journal} {\bibinfo
  {journal} {Class. Quant. Grav.}\ }\textbf {\bibinfo {volume} {37}},\ \bibinfo
  {pages} {145001} (\bibinfo {year} {2020})},\ \Eprint
  {http://arxiv.org/abs/2002.09429} {arXiv:2002.09429 [gr-qc]} \BibitemShut
  {NoStop}%
\bibitem [{\citenamefont {Messick}\ \emph {et~al.}(2017)\citenamefont {Messick}
  \emph {et~al.}}]{Messick:2016aqy}%
  \BibitemOpen
  \bibfield  {author} {\bibinfo {author} {\bibfnamefont {C.}~\bibnamefont
  {Messick}} \emph {et~al.},\ }\href {\doibase 10.1103/PhysRevD.95.042001}
  {\bibfield  {journal} {\bibinfo  {journal} {Phys. Rev. D}\ }\textbf {\bibinfo
  {volume} {95}},\ \bibinfo {pages} {042001} (\bibinfo {year} {2017})},\
  \Eprint {http://arxiv.org/abs/1604.04324} {arXiv:1604.04324 [astro-ph.IM]}
  \BibitemShut {NoStop}%
\bibitem [{\citenamefont {Sachdev}\ \emph {et~al.}(2019)\citenamefont {Sachdev}
  \emph {et~al.}}]{Sachdev:2019vvd}%
  \BibitemOpen
  \bibfield  {author} {\bibinfo {author} {\bibfnamefont {S.}~\bibnamefont
  {Sachdev}} \emph {et~al.},\ }\href@noop {} {\  (\bibinfo {year} {2019})},\
  \Eprint {http://arxiv.org/abs/1901.08580} {arXiv:1901.08580 [gr-qc]}
  \BibitemShut {NoStop}%
\bibitem [{\citenamefont {{Cannon}}\ \emph {et~al.}(2021)\citenamefont
  {{Cannon}}, \citenamefont {{Caudill}}, \citenamefont {{Chan}}, \citenamefont
  {{Cousins}}, \citenamefont {{Creighton}}, \citenamefont {{Ewing}},
  \citenamefont {{Fong}}, \citenamefont {{Godwin}}, \citenamefont {{Hanna}},
  \citenamefont {{Hooper}}, \citenamefont {{Huxford}}, \citenamefont {{Magee}},
  \citenamefont {{Meacher}}, \citenamefont {{Messick}}, \citenamefont
  {{Morisaki}}, \citenamefont {{Mukherjee}}, \citenamefont {{Ohta}},
  \citenamefont {{Pace}}, \citenamefont {{Privitera}}, \citenamefont {{de
  Ruiter}}, \citenamefont {{Sachdev}}, \citenamefont {{Singer}}, \citenamefont
  {{Singh}}, \citenamefont {{Tapia}}, \citenamefont {{Tsukada}}, \citenamefont
  {{Tsuna}}, \citenamefont {{Tsutsui}}, \citenamefont {{Ueno}}, \citenamefont
  {{Viets}}, \citenamefont {{Wade}},\ and\ \citenamefont
  {{Wade}}}]{2021SoftX..1400680C}%
  \BibitemOpen
  \bibfield  {author} {\bibinfo {author} {\bibfnamefont {K.}~\bibnamefont
  {{Cannon}}}, \bibinfo {author} {\bibfnamefont {S.}~\bibnamefont {{Caudill}}},
  \bibinfo {author} {\bibfnamefont {C.}~\bibnamefont {{Chan}}}, \bibinfo
  {author} {\bibfnamefont {B.}~\bibnamefont {{Cousins}}}, \bibinfo {author}
  {\bibfnamefont {J.~D.~E.}\ \bibnamefont {{Creighton}}}, \bibinfo {author}
  {\bibfnamefont {B.}~\bibnamefont {{Ewing}}}, \bibinfo {author} {\bibfnamefont
  {H.}~\bibnamefont {{Fong}}}, \bibinfo {author} {\bibfnamefont
  {P.}~\bibnamefont {{Godwin}}}, \bibinfo {author} {\bibfnamefont
  {C.}~\bibnamefont {{Hanna}}}, \bibinfo {author} {\bibfnamefont
  {S.}~\bibnamefont {{Hooper}}}, \bibinfo {author} {\bibfnamefont
  {R.}~\bibnamefont {{Huxford}}}, \bibinfo {author} {\bibfnamefont
  {R.}~\bibnamefont {{Magee}}}, \bibinfo {author} {\bibfnamefont
  {D.}~\bibnamefont {{Meacher}}}, \bibinfo {author} {\bibfnamefont
  {C.}~\bibnamefont {{Messick}}}, \bibinfo {author} {\bibfnamefont
  {S.}~\bibnamefont {{Morisaki}}}, \bibinfo {author} {\bibfnamefont
  {D.}~\bibnamefont {{Mukherjee}}}, \bibinfo {author} {\bibfnamefont
  {H.}~\bibnamefont {{Ohta}}}, \bibinfo {author} {\bibfnamefont
  {A.}~\bibnamefont {{Pace}}}, \bibinfo {author} {\bibfnamefont
  {S.}~\bibnamefont {{Privitera}}}, \bibinfo {author} {\bibfnamefont
  {I.}~\bibnamefont {{de Ruiter}}}, \bibinfo {author} {\bibfnamefont
  {S.}~\bibnamefont {{Sachdev}}}, \bibinfo {author} {\bibfnamefont
  {L.}~\bibnamefont {{Singer}}}, \bibinfo {author} {\bibfnamefont
  {D.}~\bibnamefont {{Singh}}}, \bibinfo {author} {\bibfnamefont
  {R.}~\bibnamefont {{Tapia}}}, \bibinfo {author} {\bibfnamefont
  {L.}~\bibnamefont {{Tsukada}}}, \bibinfo {author} {\bibfnamefont
  {D.}~\bibnamefont {{Tsuna}}}, \bibinfo {author} {\bibfnamefont
  {T.}~\bibnamefont {{Tsutsui}}}, \bibinfo {author} {\bibfnamefont
  {K.}~\bibnamefont {{Ueno}}}, \bibinfo {author} {\bibfnamefont
  {A.}~\bibnamefont {{Viets}}}, \bibinfo {author} {\bibfnamefont
  {L.}~\bibnamefont {{Wade}}}, \ and\ \bibinfo {author} {\bibfnamefont
  {M.}~\bibnamefont {{Wade}}},\ }\href {\doibase 10.1016/j.softx.2021.100680}
  {\bibfield  {journal} {\bibinfo  {journal} {SoftwareX}\ }\textbf {\bibinfo
  {volume} {14}},\ \bibinfo {eid} {100680} (\bibinfo {year} {2021})},\ \Eprint
  {http://arxiv.org/abs/2010.05082} {arXiv:2010.05082 [astro-ph.IM]}
  \BibitemShut {NoStop}%
\bibitem [{\citenamefont {LeCun}\ \emph {et~al.}(2015)\citenamefont {LeCun},
  \citenamefont {Bengio},\ and\ \citenamefont {Hinton}}]{LeCun:2015a}%
  \BibitemOpen
  \bibfield  {author} {\bibinfo {author} {\bibfnamefont {Y.}~\bibnamefont
  {LeCun}}, \bibinfo {author} {\bibfnamefont {Y.}~\bibnamefont {Bengio}}, \
  and\ \bibinfo {author} {\bibfnamefont {G.}~\bibnamefont {Hinton}},\ }\href
  {\doibase 10.1038/nature14539} {\bibfield  {journal} {\bibinfo  {journal}
  {Nature}\ }\textbf {\bibinfo {volume} {521}},\ \bibinfo {pages} {436}
  (\bibinfo {year} {2015})}\BibitemShut {NoStop}%
\bibitem [{\citenamefont {{Schmidhuber}}(2014)}]{2014arXiv1404.7828S}%
  \BibitemOpen
  \bibfield  {author} {\bibinfo {author} {\bibfnamefont {J.}~\bibnamefont
  {{Schmidhuber}}},\ }\href {\doibase 10.48550/arXiv.1404.7828} {\bibfield
  {journal} {\bibinfo  {journal} {arXiv e-prints}\ ,\ \bibinfo {eid}
  {arXiv:1404.7828}} (\bibinfo {year} {2014})},\ \Eprint
  {http://arxiv.org/abs/1404.7828} {arXiv:1404.7828 [cs.NE]} \BibitemShut
  {NoStop}%
\bibitem [{\citenamefont {Cannon}\ \emph {et~al.}(2010)\citenamefont {Cannon},
  \citenamefont {Chapman}, \citenamefont {Hanna}, \citenamefont {Keppel},
  \citenamefont {Searle},\ and\ \citenamefont {Weinstein}}]{Cannon:2010qh}%
  \BibitemOpen
  \bibfield  {author} {\bibinfo {author} {\bibfnamefont {K.}~\bibnamefont
  {Cannon}}, \bibinfo {author} {\bibfnamefont {A.}~\bibnamefont {Chapman}},
  \bibinfo {author} {\bibfnamefont {C.}~\bibnamefont {Hanna}}, \bibinfo
  {author} {\bibfnamefont {D.}~\bibnamefont {Keppel}}, \bibinfo {author}
  {\bibfnamefont {A.~C.}\ \bibnamefont {Searle}}, \ and\ \bibinfo {author}
  {\bibfnamefont {A.~J.}\ \bibnamefont {Weinstein}},\ }\href {\doibase
  10.1103/PhysRevD.82.044025} {\bibfield  {journal} {\bibinfo  {journal} {Phys.
  Rev. D}\ }\textbf {\bibinfo {volume} {82}},\ \bibinfo {pages} {044025}
  (\bibinfo {year} {2010})},\ \Eprint {http://arxiv.org/abs/1005.0012}
  {arXiv:1005.0012 [gr-qc]} \BibitemShut {NoStop}%
\bibitem [{\citenamefont {Cannon}\ \emph
  {et~al.}(2011{\natexlab{a}})\citenamefont {Cannon}, \citenamefont {Hanna},\
  and\ \citenamefont {Keppel}}]{Cannon:2011xk}%
  \BibitemOpen
  \bibfield  {author} {\bibinfo {author} {\bibfnamefont {K.}~\bibnamefont
  {Cannon}}, \bibinfo {author} {\bibfnamefont {C.}~\bibnamefont {Hanna}}, \
  and\ \bibinfo {author} {\bibfnamefont {D.}~\bibnamefont {Keppel}},\ }\href
  {\doibase 10.1103/PhysRevD.84.084003} {\bibfield  {journal} {\bibinfo
  {journal} {Phys. Rev. D}\ }\textbf {\bibinfo {volume} {84}},\ \bibinfo
  {pages} {084003} (\bibinfo {year} {2011}{\natexlab{a}})},\ \Eprint
  {http://arxiv.org/abs/1101.4939} {arXiv:1101.4939 [gr-qc]} \BibitemShut
  {NoStop}%
\bibitem [{\citenamefont {Cannon}\ \emph
  {et~al.}(2012{\natexlab{a}})\citenamefont {Cannon}, \citenamefont {Hanna},\
  and\ \citenamefont {Keppel}}]{Cannon:2011rj}%
  \BibitemOpen
  \bibfield  {author} {\bibinfo {author} {\bibfnamefont {K.}~\bibnamefont
  {Cannon}}, \bibinfo {author} {\bibfnamefont {C.}~\bibnamefont {Hanna}}, \
  and\ \bibinfo {author} {\bibfnamefont {D.}~\bibnamefont {Keppel}},\ }\href
  {\doibase 10.1103/PhysRevD.85.081504} {\bibfield  {journal} {\bibinfo
  {journal} {Phys. Rev. D}\ }\textbf {\bibinfo {volume} {85}},\ \bibinfo
  {pages} {081504} (\bibinfo {year} {2012}{\natexlab{a}})},\ \Eprint
  {http://arxiv.org/abs/1108.5618} {arXiv:1108.5618 [gr-qc]} \BibitemShut
  {NoStop}%
\bibitem [{\citenamefont {Cannon}\ \emph
  {et~al.}(2011{\natexlab{b}})\citenamefont {Cannon}, \citenamefont {Hanna},
  \citenamefont {Keppel},\ and\ \citenamefont {Searle}}]{Cannon:2011tb}%
  \BibitemOpen
  \bibfield  {author} {\bibinfo {author} {\bibfnamefont {K.}~\bibnamefont
  {Cannon}}, \bibinfo {author} {\bibfnamefont {C.}~\bibnamefont {Hanna}},
  \bibinfo {author} {\bibfnamefont {D.}~\bibnamefont {Keppel}}, \ and\ \bibinfo
  {author} {\bibfnamefont {A.~C.}\ \bibnamefont {Searle}},\ }\href {\doibase
  10.1103/PhysRevD.83.084053} {\bibfield  {journal} {\bibinfo  {journal} {Phys.
  Rev. D}\ }\textbf {\bibinfo {volume} {83}},\ \bibinfo {pages} {084053}
  (\bibinfo {year} {2011}{\natexlab{b}})},\ \Eprint
  {http://arxiv.org/abs/1101.0584} {arXiv:1101.0584 [physics.data-an]}
  \BibitemShut {NoStop}%
\bibitem [{\citenamefont {Soni}\ \emph {et~al.}(2020)\citenamefont {Soni} \emph
  {et~al.}}]{LIGO:2020zwl}%
  \BibitemOpen
  \bibfield  {author} {\bibinfo {author} {\bibfnamefont {S.}~\bibnamefont
  {Soni}} \emph {et~al.} (\bibinfo {collaboration} {LIGO}),\ }\href {\doibase
  10.1088/1361-6382/abc906} {\bibfield  {journal} {\bibinfo  {journal} {Class.
  Quant. Grav.}\ }\textbf {\bibinfo {volume} {38}},\ \bibinfo {pages} {025016}
  (\bibinfo {year} {2020})},\ \Eprint {http://arxiv.org/abs/2007.14876}
  {arXiv:2007.14876 [astro-ph.IM]} \BibitemShut {NoStop}%
\bibitem [{\citenamefont {Powell}\ \emph {et~al.}(2023)\citenamefont {Powell},
  \citenamefont {Sun}, \citenamefont {Gereb}, \citenamefont {Lasky},\ and\
  \citenamefont {Dollmann}}]{Powell:2022pcg}%
  \BibitemOpen
  \bibfield  {author} {\bibinfo {author} {\bibfnamefont {J.}~\bibnamefont
  {Powell}}, \bibinfo {author} {\bibfnamefont {L.}~\bibnamefont {Sun}},
  \bibinfo {author} {\bibfnamefont {K.}~\bibnamefont {Gereb}}, \bibinfo
  {author} {\bibfnamefont {P.~D.}\ \bibnamefont {Lasky}}, \ and\ \bibinfo
  {author} {\bibfnamefont {M.}~\bibnamefont {Dollmann}},\ }\href {\doibase
  10.1088/1361-6382/acb038} {\bibfield  {journal} {\bibinfo  {journal} {Class.
  Quant. Grav.}\ }\textbf {\bibinfo {volume} {40}},\ \bibinfo {pages} {035006}
  (\bibinfo {year} {2023})},\ \Eprint {http://arxiv.org/abs/2207.00207}
  {arXiv:2207.00207 [astro-ph.IM]} \BibitemShut {NoStop}%
\bibitem [{\citenamefont {{Goodfellow}}\ \emph {et~al.}(2014)\citenamefont
  {{Goodfellow}}, \citenamefont {{Pouget-Abadie}}, \citenamefont {{Mirza}},
  \citenamefont {{Xu}}, \citenamefont {{Warde-Farley}}, \citenamefont
  {{Ozair}}, \citenamefont {{Courville}},\ and\ \citenamefont
  {{Bengio}}}]{2014arXiv1406.2661G}%
  \BibitemOpen
  \bibfield  {author} {\bibinfo {author} {\bibfnamefont {I.~J.}\ \bibnamefont
  {{Goodfellow}}}, \bibinfo {author} {\bibfnamefont {J.}~\bibnamefont
  {{Pouget-Abadie}}}, \bibinfo {author} {\bibfnamefont {M.}~\bibnamefont
  {{Mirza}}}, \bibinfo {author} {\bibfnamefont {B.}~\bibnamefont {{Xu}}},
  \bibinfo {author} {\bibfnamefont {D.}~\bibnamefont {{Warde-Farley}}},
  \bibinfo {author} {\bibfnamefont {S.}~\bibnamefont {{Ozair}}}, \bibinfo
  {author} {\bibfnamefont {A.}~\bibnamefont {{Courville}}}, \ and\ \bibinfo
  {author} {\bibfnamefont {Y.}~\bibnamefont {{Bengio}}},\ }\href {\doibase
  10.48550/arXiv.1406.2661} {\bibfield  {journal} {\bibinfo  {journal} {arXiv
  e-prints}\ ,\ \bibinfo {eid} {arXiv:1406.2661}} (\bibinfo {year} {2014})},\
  \Eprint {http://arxiv.org/abs/1406.2661} {arXiv:1406.2661 [stat.ML]}
  \BibitemShut {NoStop}%
\bibitem [{\citenamefont {Accadia}\ \emph {et~al.}(2010)\citenamefont {Accadia}
  \emph {et~al.}}]{Accadia:2010zzb}%
  \BibitemOpen
  \bibfield  {author} {\bibinfo {author} {\bibfnamefont {T.}~\bibnamefont
  {Accadia}} \emph {et~al.},\ }\href {\doibase 10.1088/0264-9381/27/19/194011}
  {\bibfield  {journal} {\bibinfo  {journal} {Class. Quant. Grav.}\ }\textbf
  {\bibinfo {volume} {27}},\ \bibinfo {pages} {194011} (\bibinfo {year}
  {2010})}\BibitemShut {NoStop}%
\bibitem [{\citenamefont {Soni}\ \emph {et~al.}(2024)\citenamefont {Soni} \emph
  {et~al.}}]{Soni:2024isj}%
  \BibitemOpen
  \bibfield  {author} {\bibinfo {author} {\bibfnamefont {S.}~\bibnamefont
  {Soni}} \emph {et~al.},\ }\href@noop {} {\  (\bibinfo {year} {2024})},\
  \Eprint {http://arxiv.org/abs/2409.02831} {arXiv:2409.02831 [astro-ph.IM]}
  \BibitemShut {NoStop}%
\bibitem [{\citenamefont {Udall}\ and\ \citenamefont
  {Davis}(2023)}]{Udall:2022vkv}%
  \BibitemOpen
  \bibfield  {author} {\bibinfo {author} {\bibfnamefont {R.}~\bibnamefont
  {Udall}}\ and\ \bibinfo {author} {\bibfnamefont {D.}~\bibnamefont {Davis}},\
  }\href {\doibase 10.1063/5.0136896} {\bibfield  {journal} {\bibinfo
  {journal} {Appl. Phys. Lett.}\ }\textbf {\bibinfo {volume} {122}},\ \bibinfo
  {pages} {094103} (\bibinfo {year} {2023})},\ \Eprint
  {http://arxiv.org/abs/2211.15867} {arXiv:2211.15867 [astro-ph.IM]}
  \BibitemShut {NoStop}%
\bibitem [{\citenamefont {Owen}\ and\ \citenamefont
  {Sathyaprakash}(1999)}]{Owen:1998dk}%
  \BibitemOpen
  \bibfield  {author} {\bibinfo {author} {\bibfnamefont {B.~J.}\ \bibnamefont
  {Owen}}\ and\ \bibinfo {author} {\bibfnamefont {B.~S.}\ \bibnamefont
  {Sathyaprakash}},\ }\href {\doibase 10.1103/PhysRevD.60.022002} {\bibfield
  {journal} {\bibinfo  {journal} {Phys. Rev. D}\ }\textbf {\bibinfo {volume}
  {60}},\ \bibinfo {pages} {022002} (\bibinfo {year} {1999})},\ \Eprint
  {http://arxiv.org/abs/gr-qc/9808076} {arXiv:gr-qc/9808076} \BibitemShut
  {NoStop}%
\bibitem [{\citenamefont {Harry}\ \emph {et~al.}(2009)\citenamefont {Harry},
  \citenamefont {Allen},\ and\ \citenamefont {Sathyaprakash}}]{Harry:2009ea}%
  \BibitemOpen
  \bibfield  {author} {\bibinfo {author} {\bibfnamefont {I.~W.}\ \bibnamefont
  {Harry}}, \bibinfo {author} {\bibfnamefont {B.}~\bibnamefont {Allen}}, \ and\
  \bibinfo {author} {\bibfnamefont {B.~S.}\ \bibnamefont {Sathyaprakash}},\
  }\href {\doibase 10.1103/PhysRevD.80.104014} {\bibfield  {journal} {\bibinfo
  {journal} {Phys. Rev. D}\ }\textbf {\bibinfo {volume} {80}},\ \bibinfo
  {pages} {104014} (\bibinfo {year} {2009})},\ \Eprint
  {http://arxiv.org/abs/0908.2090} {arXiv:0908.2090 [gr-qc]} \BibitemShut
  {NoStop}%
\bibitem [{\citenamefont {Privitera}\ \emph {et~al.}(2014)\citenamefont
  {Privitera}, \citenamefont {Mohapatra}, \citenamefont {Ajith}, \citenamefont
  {Cannon}, \citenamefont {Fotopoulos}, \citenamefont {Frei}, \citenamefont
  {Hanna}, \citenamefont {Weinstein},\ and\ \citenamefont
  {Whelan}}]{Privitera:2013xza}%
  \BibitemOpen
  \bibfield  {author} {\bibinfo {author} {\bibfnamefont {S.}~\bibnamefont
  {Privitera}}, \bibinfo {author} {\bibfnamefont {S.~R.~P.}\ \bibnamefont
  {Mohapatra}}, \bibinfo {author} {\bibfnamefont {P.}~\bibnamefont {Ajith}},
  \bibinfo {author} {\bibfnamefont {K.}~\bibnamefont {Cannon}}, \bibinfo
  {author} {\bibfnamefont {N.}~\bibnamefont {Fotopoulos}}, \bibinfo {author}
  {\bibfnamefont {M.~A.}\ \bibnamefont {Frei}}, \bibinfo {author}
  {\bibfnamefont {C.}~\bibnamefont {Hanna}}, \bibinfo {author} {\bibfnamefont
  {A.~J.}\ \bibnamefont {Weinstein}}, \ and\ \bibinfo {author} {\bibfnamefont
  {J.~T.}\ \bibnamefont {Whelan}},\ }\href {\doibase
  10.1103/PhysRevD.89.024003} {\bibfield  {journal} {\bibinfo  {journal} {Phys.
  Rev. D}\ }\textbf {\bibinfo {volume} {89}},\ \bibinfo {pages} {024003}
  (\bibinfo {year} {2014})},\ \Eprint {http://arxiv.org/abs/1310.5633}
  {arXiv:1310.5633 [gr-qc]} \BibitemShut {NoStop}%
\bibitem [{\citenamefont {Khan}\ \emph {et~al.}(2016)\citenamefont {Khan},
  \citenamefont {Husa}, \citenamefont {Hannam}, \citenamefont {Ohme},
  \citenamefont {P\"urrer}, \citenamefont {Jim\'enez~Forteza},\ and\
  \citenamefont {Boh\'e}}]{Khan:2015jqa}%
  \BibitemOpen
  \bibfield  {author} {\bibinfo {author} {\bibfnamefont {S.}~\bibnamefont
  {Khan}}, \bibinfo {author} {\bibfnamefont {S.}~\bibnamefont {Husa}}, \bibinfo
  {author} {\bibfnamefont {M.}~\bibnamefont {Hannam}}, \bibinfo {author}
  {\bibfnamefont {F.}~\bibnamefont {Ohme}}, \bibinfo {author} {\bibfnamefont
  {M.}~\bibnamefont {P\"urrer}}, \bibinfo {author} {\bibfnamefont
  {X.}~\bibnamefont {Jim\'enez~Forteza}}, \ and\ \bibinfo {author}
  {\bibfnamefont {A.}~\bibnamefont {Boh\'e}},\ }\href {\doibase
  10.1103/PhysRevD.93.044007} {\bibfield  {journal} {\bibinfo  {journal} {Phys.
  Rev. D}\ }\textbf {\bibinfo {volume} {93}},\ \bibinfo {pages} {044007}
  (\bibinfo {year} {2016})},\ \Eprint {http://arxiv.org/abs/1508.07253}
  {arXiv:1508.07253 [gr-qc]} \BibitemShut {NoStop}%
\bibitem [{\citenamefont {Sakon}\ \emph {et~al.}(2022)\citenamefont {Sakon}
  \emph {et~al.}}]{Sakon:2022ibh}%
  \BibitemOpen
  \bibfield  {author} {\bibinfo {author} {\bibfnamefont {S.}~\bibnamefont
  {Sakon}} \emph {et~al.},\ }\href@noop {} {\  (\bibinfo {year} {2022})},\
  \Eprint {http://arxiv.org/abs/2211.16674} {arXiv:2211.16674 [gr-qc]}
  \BibitemShut {NoStop}%
\bibitem [{\citenamefont {Boh\'e}\ \emph {et~al.}(2017)\citenamefont {Boh\'e}
  \emph {et~al.}}]{Bohe:2016gbl}%
  \BibitemOpen
  \bibfield  {author} {\bibinfo {author} {\bibfnamefont {A.}~\bibnamefont
  {Boh\'e}} \emph {et~al.},\ }\href {\doibase 10.1103/PhysRevD.95.044028}
  {\bibfield  {journal} {\bibinfo  {journal} {Phys. Rev. D}\ }\textbf {\bibinfo
  {volume} {95}},\ \bibinfo {pages} {044028} (\bibinfo {year} {2017})},\
  \Eprint {http://arxiv.org/abs/1611.03703} {arXiv:1611.03703 [gr-qc]}
  \BibitemShut {NoStop}%
\bibitem [{\citenamefont {Cannon}\ \emph
  {et~al.}(2012{\natexlab{b}})\citenamefont {Cannon} \emph
  {et~al.}}]{Cannon:2011vi}%
  \BibitemOpen
  \bibfield  {author} {\bibinfo {author} {\bibfnamefont {K.}~\bibnamefont
  {Cannon}} \emph {et~al.},\ }\href {\doibase 10.1088/0004-637X/748/2/136}
  {\bibfield  {journal} {\bibinfo  {journal} {Astrophys. J.}\ }\textbf
  {\bibinfo {volume} {748}},\ \bibinfo {pages} {136} (\bibinfo {year}
  {2012}{\natexlab{b}})},\ \Eprint {http://arxiv.org/abs/1107.2665}
  {arXiv:1107.2665 [astro-ph.IM]} \BibitemShut {NoStop}%
\bibitem [{\citenamefont {Sachdev}\ \emph {et~al.}(2020)\citenamefont {Sachdev}
  \emph {et~al.}}]{Sachdev:2020lfd}%
  \BibitemOpen
  \bibfield  {author} {\bibinfo {author} {\bibfnamefont {S.}~\bibnamefont
  {Sachdev}} \emph {et~al.},\ }\href {\doibase 10.3847/2041-8213/abc753}
  {\bibfield  {journal} {\bibinfo  {journal} {Astrophys. J. Lett.}\ }\textbf
  {\bibinfo {volume} {905}},\ \bibinfo {pages} {L25} (\bibinfo {year}
  {2020})},\ \Eprint {http://arxiv.org/abs/2008.04288} {arXiv:2008.04288
  [astro-ph.HE]} \BibitemShut {NoStop}%
\bibitem [{\citenamefont {Abadie}\ \emph {et~al.}(2012)\citenamefont {Abadie}
  \emph {et~al.}}]{LIGOScientific:2011jth}%
  \BibitemOpen
  \bibfield  {author} {\bibinfo {author} {\bibfnamefont {J.}~\bibnamefont
  {Abadie}} \emph {et~al.} (\bibinfo {collaboration} {LIGO Scientific,
  VIRGO}),\ }\href {\doibase 10.1103/PhysRevD.85.082002} {\bibfield  {journal}
  {\bibinfo  {journal} {Phys. Rev. D}\ }\textbf {\bibinfo {volume} {85}},\
  \bibinfo {pages} {082002} (\bibinfo {year} {2012})},\ \Eprint
  {http://arxiv.org/abs/1111.7314} {arXiv:1111.7314 [gr-qc]} \BibitemShut
  {NoStop}%
\bibitem [{\citenamefont {Babak}\ \emph {et~al.}(2013)\citenamefont {Babak},
  \citenamefont {Biswas}, \citenamefont {Brady}, \citenamefont {Brown},
  \citenamefont {Cannon} \emph {et~al.}}]{Babak:2012zx}%
  \BibitemOpen
  \bibfield  {author} {\bibinfo {author} {\bibfnamefont {S.}~\bibnamefont
  {Babak}}, \bibinfo {author} {\bibfnamefont {R.}~\bibnamefont {Biswas}},
  \bibinfo {author} {\bibfnamefont {P.}~\bibnamefont {Brady}}, \bibinfo
  {author} {\bibfnamefont {D.}~\bibnamefont {Brown}}, \bibinfo {author}
  {\bibfnamefont {K.}~\bibnamefont {Cannon}},  \emph {et~al.},\ }\href
  {\doibase 10.1103/PhysRevD.87.024033} {\bibfield  {journal} {\bibinfo
  {journal} {Phys. Rev. D}\ }\textbf {\bibinfo {volume} {87}},\ \bibinfo
  {pages} {024033} (\bibinfo {year} {2013})},\ \Eprint
  {http://arxiv.org/abs/1208.3491} {arXiv:1208.3491 [gr-qc]} \BibitemShut
  {NoStop}%
\bibitem [{\citenamefont {{Xu}}\ \emph {et~al.}(2015)\citenamefont {{Xu}},
  \citenamefont {{Wang}}, \citenamefont {{Chen}},\ and\ \citenamefont
  {{Li}}}]{2015arXiv150500853X}%
  \BibitemOpen
  \bibfield  {author} {\bibinfo {author} {\bibfnamefont {B.}~\bibnamefont
  {{Xu}}}, \bibinfo {author} {\bibfnamefont {N.}~\bibnamefont {{Wang}}},
  \bibinfo {author} {\bibfnamefont {T.}~\bibnamefont {{Chen}}}, \ and\ \bibinfo
  {author} {\bibfnamefont {M.}~\bibnamefont {{Li}}},\ }\href {\doibase
  10.48550/arXiv.1505.00853} {\bibfield  {journal} {\bibinfo  {journal} {arXiv
  e-prints}\ ,\ \bibinfo {eid} {arXiv:1505.00853}} (\bibinfo {year} {2015})},\
  \Eprint {http://arxiv.org/abs/1505.00853} {arXiv:1505.00853 [cs.LG]}
  \BibitemShut {NoStop}%
\bibitem [{\citenamefont {Macas}\ \emph {et~al.}(2022)\citenamefont {Macas},
  \citenamefont {Pooley}, \citenamefont {Nuttall}, \citenamefont {Davis},
  \citenamefont {Dyer}, \citenamefont {Lecoeuche}, \citenamefont {Lyman},
  \citenamefont {McIver},\ and\ \citenamefont {Rink}}]{Macas:2022afm}%
  \BibitemOpen
  \bibfield  {author} {\bibinfo {author} {\bibfnamefont {R.}~\bibnamefont
  {Macas}}, \bibinfo {author} {\bibfnamefont {J.}~\bibnamefont {Pooley}},
  \bibinfo {author} {\bibfnamefont {L.~K.}\ \bibnamefont {Nuttall}}, \bibinfo
  {author} {\bibfnamefont {D.}~\bibnamefont {Davis}}, \bibinfo {author}
  {\bibfnamefont {M.~J.}\ \bibnamefont {Dyer}}, \bibinfo {author}
  {\bibfnamefont {Y.}~\bibnamefont {Lecoeuche}}, \bibinfo {author}
  {\bibfnamefont {J.~D.}\ \bibnamefont {Lyman}}, \bibinfo {author}
  {\bibfnamefont {J.}~\bibnamefont {McIver}}, \ and\ \bibinfo {author}
  {\bibfnamefont {K.}~\bibnamefont {Rink}},\ }\href {\doibase
  10.1103/PhysRevD.105.103021} {\bibfield  {journal} {\bibinfo  {journal}
  {Phys. Rev. D}\ }\textbf {\bibinfo {volume} {105}},\ \bibinfo {pages}
  {103021} (\bibinfo {year} {2022})},\ \Eprint
  {http://arxiv.org/abs/2202.00344} {arXiv:2202.00344 [astro-ph.HE]}
  \BibitemShut {NoStop}%
\bibitem [{\citenamefont {Kingma}\ and\ \citenamefont
  {Ba}(2014)}]{Kingma:2014vow}%
  \BibitemOpen
  \bibfield  {author} {\bibinfo {author} {\bibfnamefont {D.~P.}\ \bibnamefont
  {Kingma}}\ and\ \bibinfo {author} {\bibfnamefont {J.}~\bibnamefont {Ba}}\
  }(\bibinfo {year} {2014})\ \Eprint {http://arxiv.org/abs/1412.6980}
  {arXiv:1412.6980 [cs.LG]} \BibitemShut {NoStop}%
\bibitem [{\citenamefont {Van~Rijsbergen}(1979)}]{van1979information}%
  \BibitemOpen
  \bibfield  {author} {\bibinfo {author} {\bibfnamefont {C.~J.}\ \bibnamefont
  {Van~Rijsbergen}},\ }\href@noop {} {\enquote {\bibinfo {title} {Information
  retrieval. 2nd. newton, ma},}\ } (\bibinfo {year} {1979})\BibitemShut
  {NoStop}%
\bibitem [{\citenamefont {Youden}(1950)}]{Youden1950}%
  \BibitemOpen
  \bibfield  {author} {\bibinfo {author} {\bibfnamefont {W.~J.}\ \bibnamefont
  {Youden}},\ }\href {\doibase
  https://doi.org/10.1002/1097-0142(1950)3:1<32::AID-CNCR2820030106>3.0.CO;2-3}
  {\bibfield  {journal} {\bibinfo  {journal} {Cancer}\ }\textbf {\bibinfo
  {volume} {3}},\ \bibinfo {pages} {32} (\bibinfo {year} {1950})},\ \Eprint
  {http://arxiv.org/abs/https://doi.org/10.1002/1097-0142(1950)3:1<32::AID-CNCR2820030106>3.0.CO;2-3}
  {https://doi.org/10.1002/1097-0142(1950)3:1<32::AID-CNCR2820030106>3.0.CO;2-3}
  \BibitemShut {NoStop}%
\bibitem [{\citenamefont {Matthews}(1975)}]{MATTHEWS1975442}%
  \BibitemOpen
  \bibfield  {author} {\bibinfo {author} {\bibfnamefont {B.}~\bibnamefont
  {Matthews}},\ }\href {\doibase https://doi.org/10.1016/0005-2795(75)90109-9}
  {\bibfield  {journal} {\bibinfo  {journal} {Biochimica et Biophysica Acta
  (BBA) - Protein Structure}\ }\textbf {\bibinfo {volume} {405}},\ \bibinfo
  {pages} {442} (\bibinfo {year} {1975})}\BibitemShut {NoStop}%
\bibitem [{\citenamefont {Cohen}(1960)}]{doi:10.1177/001316446002000104}%
  \BibitemOpen
  \bibfield  {author} {\bibinfo {author} {\bibfnamefont {J.}~\bibnamefont
  {Cohen}},\ }\href {\doibase 10.1177/001316446002000104} {\bibfield  {journal}
  {\bibinfo  {journal} {Educational and Psychological Measurement}\ }\textbf
  {\bibinfo {volume} {20}},\ \bibinfo {pages} {37} (\bibinfo {year} {1960})},\
  \Eprint {http://arxiv.org/abs/https://doi.org/10.1177/001316446002000104}
  {https://doi.org/10.1177/001316446002000104} \BibitemShut {NoStop}%
\bibitem [{\citenamefont {Stehman}(1997)}]{STEHMAN199777}%
  \BibitemOpen
  \bibfield  {author} {\bibinfo {author} {\bibfnamefont {S.~V.}\ \bibnamefont
  {Stehman}},\ }\href {\doibase https://doi.org/10.1016/S0034-4257(97)00083-7}
  {\bibfield  {journal} {\bibinfo  {journal} {Remote Sensing of Environment}\
  }\textbf {\bibinfo {volume} {62}},\ \bibinfo {pages} {77} (\bibinfo {year}
  {1997})}\BibitemShut {NoStop}%
\bibitem [{\citenamefont {Woosley}(2017)}]{Woosley:2016hmi}%
  \BibitemOpen
  \bibfield  {author} {\bibinfo {author} {\bibfnamefont {S.~E.}\ \bibnamefont
  {Woosley}},\ }\href {\doibase 10.3847/1538-4357/836/2/244} {\bibfield
  {journal} {\bibinfo  {journal} {Astrophys. J.}\ }\textbf {\bibinfo {volume}
  {836}},\ \bibinfo {pages} {244} (\bibinfo {year} {2017})},\ \Eprint
  {http://arxiv.org/abs/1608.08939} {arXiv:1608.08939 [astro-ph.HE]}
  \BibitemShut {NoStop}%
\bibitem [{\citenamefont {Belczynski}\ \emph {et~al.}(2016)\citenamefont
  {Belczynski} \emph {et~al.}}]{Belczynski:2016jno}%
  \BibitemOpen
  \bibfield  {author} {\bibinfo {author} {\bibfnamefont {K.}~\bibnamefont
  {Belczynski}} \emph {et~al.},\ }\href {\doibase 10.1051/0004-6361/201628980}
  {\bibfield  {journal} {\bibinfo  {journal} {Astron. Astrophys.}\ }\textbf
  {\bibinfo {volume} {594}},\ \bibinfo {pages} {A97} (\bibinfo {year}
  {2016})},\ \Eprint {http://arxiv.org/abs/1607.03116} {arXiv:1607.03116
  [astro-ph.HE]} \BibitemShut {NoStop}%
\bibitem [{\citenamefont {Giacobbo}\ \emph {et~al.}(2018)\citenamefont
  {Giacobbo}, \citenamefont {Mapelli},\ and\ \citenamefont
  {Spera}}]{Giacobbo:2017qhh}%
  \BibitemOpen
  \bibfield  {author} {\bibinfo {author} {\bibfnamefont {N.}~\bibnamefont
  {Giacobbo}}, \bibinfo {author} {\bibfnamefont {M.}~\bibnamefont {Mapelli}}, \
  and\ \bibinfo {author} {\bibfnamefont {M.}~\bibnamefont {Spera}},\ }\href
  {\doibase 10.1093/mnras/stx2933} {\bibfield  {journal} {\bibinfo  {journal}
  {Mon. Not. Roy. Astron. Soc.}\ }\textbf {\bibinfo {volume} {474}},\ \bibinfo
  {pages} {2959} (\bibinfo {year} {2018})},\ \Eprint
  {http://arxiv.org/abs/1711.03556} {arXiv:1711.03556 [astro-ph.SR]}
  \BibitemShut {NoStop}%
\bibitem [{\citenamefont {Spera}\ and\ \citenamefont
  {Mapelli}(2017)}]{Spera:2017fyx}%
  \BibitemOpen
  \bibfield  {author} {\bibinfo {author} {\bibfnamefont {M.}~\bibnamefont
  {Spera}}\ and\ \bibinfo {author} {\bibfnamefont {M.}~\bibnamefont
  {Mapelli}},\ }\href {\doibase 10.1093/mnras/stx1576} {\bibfield  {journal}
  {\bibinfo  {journal} {Mon. Not. Roy. Astron. Soc.}\ }\textbf {\bibinfo
  {volume} {470}},\ \bibinfo {pages} {4739} (\bibinfo {year} {2017})},\ \Eprint
  {http://arxiv.org/abs/1706.06109} {arXiv:1706.06109 [astro-ph.SR]}
  \BibitemShut {NoStop}%
\bibitem [{\citenamefont {Abbott}\ \emph
  {et~al.}(2020{\natexlab{d}})\citenamefont {Abbott} \emph
  {et~al.}}]{LIGOScientific:2020ufj}%
  \BibitemOpen
  \bibfield  {author} {\bibinfo {author} {\bibfnamefont {R.}~\bibnamefont
  {Abbott}} \emph {et~al.} (\bibinfo {collaboration} {LIGO Scientific,
  Virgo}),\ }\href {\doibase 10.3847/2041-8213/aba493} {\bibfield  {journal}
  {\bibinfo  {journal} {Astrophys. J. Lett.}\ }\textbf {\bibinfo {volume}
  {900}},\ \bibinfo {pages} {L13} (\bibinfo {year} {2020}{\natexlab{d}})},\
  \Eprint {http://arxiv.org/abs/2009.01190} {arXiv:2009.01190 [astro-ph.HE]}
  \BibitemShut {NoStop}%
\bibitem [{\citenamefont {{LIGO Scientific Collaboration and Virgo
  Collaboration}}(2018)}]{lalsuite}%
  \BibitemOpen
  \bibfield  {author} {\bibinfo {author} {\bibnamefont {{LIGO Scientific
  Collaboration and Virgo Collaboration}}},\ }\href {\doibase
  10.7935/GT1W-FZ16} {\enquote {\bibinfo {title} {{LALSuite software}},}\ }
  (\bibinfo {year} {2018})\BibitemShut {NoStop}%
\end{thebibliography}%
\end{document}